\documentclass[a4paper,11pt]{article}
\pdfoutput=1 % if your are submitting a pdflatex (i.e. if you have images in pdf, png or jpg format)

\usepackage{jcappub} % for details on the use of the package, please see the JCAP-author-manual

\usepackage[T1]{fontenc} 
\usepackage{verbatim}
\usepackage[utf8]{inputenc}
\usepackage[american]{babel}
\usepackage{graphicx}
\usepackage{epsfig}
\usepackage{adjustbox}
\usepackage{booktabs}
\usepackage{multirow}
\usepackage{dcolumn}
\usepackage{amsmath}
\usepackage{amsfonts}
\usepackage{mathtools}
\usepackage{amssymb}
\usepackage{empheq}
\usepackage{subcaption} %--> for subfigures: rarely used anyway...
\usepackage{caption}
\usepackage{epstopdf}
\usepackage{bm}
\usepackage{bbm}
\usepackage{dsfont}
\usepackage{mathrsfs}
\usepackage{version}
\usepackage{enumerate}
\usepackage{braket}
\usepackage{enumitem}
\usepackage{transparent}
\usepackage{pifont}
\usepackage{colortbl}
\usepackage{rotating}
\usepackage{adjustbox}
\usepackage{cancel}
\usepackage{tabularx}
\usepackage{soul}
\usepackage{pdfpages}
\usepackage{microtype}
\usepackage{ulem}
\usepackage[table]{xcolor}
\usepackage{siunitx}
\usepackage{float} %--> can use H positioning...
\usepackage[nocompress]{cite} 
\usepackage{marginnote}
\usepackage{relsize}
\usepackage{feynmp-auto}

\usepackage{hyperref}
\hypersetup{colorlinks=true,%linkcolor=royalblue,citecolor=magenta,
linkcolor=navyblue,citecolor=coralred,urlcolor=green(munsell),pdfencoding=auto}
\definecolor{navyblue}{rgb}{0.0, 0.0, 0.5}
\definecolor{bleudefrance}{rgb}{0.19, 0.55, 0.91}
\definecolor{coralred}{rgb}{1.0, 0.25, 0.25}
\definecolor{royalblue}{rgb}{0.25, 0.41, 0.88}
\definecolor{cadmiumgreen}{rgb}{0.0, 0.42, 0.24}
\definecolor{green(munsell)}{rgb}{0.0, 0.66, 0.47}
\definecolor{blue-violet}{rgb}{0.54, 0.17, 0.89}
\definecolor{darkviolet}{rgb}{0.58, 0.0, 0.83}
\definecolor{orange(colorwheel)}{rgb}{1.0, 0.5, 0.0}
\definecolor{internationalorange}{rgb}{1.0, 0.31, 0.0}

\definecolor{magenta(process)}{rgb}{1.0, 0.0, 0.56}

\definecolor{darkspringgreen}{rgb}{0.09, 0.45, 0.27}

\definecolor{royalblue(web)}{rgb}{0.25, 0.41, 0.88}

\definecolor{cadmiumorange}{rgb}{0.93, 0.53, 0.18}

\definecolor{heliotrope}{rgb}{0.87, 0.45, 1.0}

\makeatletter
\renewcommand*{\@textcolor}[3]{%
\protect\leavevmode
\begingroup
\color#1{#2}#3%
\endgroup
}
\makeatother

\makeatletter
\newlength{\apb@width}
\newcommand{\autoparbox}[2][c]{\settowidth{\apb@width}{#2}\parbox[#1]{\apb@width}{#2}}

\makeatother

\DeclareCaptionLabelSeparator{colquad}{.\,\,\,}%{:\quad}
\DeclarePairedDelimiter{\abs}{\lvert}{\rvert}

\makeatletter
\let\save@mathaccent\mathaccent
\newcommand*\if@single[3]{%
\setbox0\hbox{${\mathaccent"0362{#1}}^H$}%
\setbox2\hbox{${\mathaccent"0362{\kern0pt#1}}^H$}%
\ifdim\ht0=\ht2 #3\else #2\fi
}
\newcommand*\rel@kern[1]{\kern#1\dimexpr\macc@kerna}
\newcommand*\widebar[1]{\@ifnextchar^{{\wide@bar{#1}{0}}}{\wide@bar{#1}{1}}}
\newcommand*\wide@bar[2]{\if@single{#1}{\wide@bar@{#1}{#2}{1}}{\wide@bar@{#1}{#2}{2}}}
\newcommand*\wide@bar@[3]{%
\begingroup
\def\mathaccent##1##2{%
\let\mathaccent\save@mathaccent
\if#32 \let\macc@nucleus\first@char \fi
\setbox\z@\hbox{$\macc@style{\macc@nucleus}_{}$}%
\setbox\tw@\hbox{$\macc@style{\macc@nucleus}{}_{}$}%
\dimen@\wd\tw@
\advance\dimen@-\wd\z@
\divide\dimen@ 3
\@tempdima\wd\tw@
\advance\@tempdima-\scriptspace
\divide\@tempdima 10
\advance\dimen@-\@tempdima
\ifdim\dimen@>\z@ \dimen@0pt\fi
\rel@kern{0.6}\kern-\dimen@
\if#31
\overline{\rel@kern{-0.6}\kern\dimen@\macc@nucleus\rel@kern{0.4}\kern\dimen@}%
\advance\dimen@0.4\dimexpr\macc@kerna
\let\final@kern#2%
\ifdim\dimen@<\z@ \let\final@kern1\fi
\if\final@kern1 \kern-\dimen@\fi
\else
\overline{\rel@kern{-0.6}\kern\dimen@#1}%
\fi
}%
\macc@depth\@ne
\let\math@bgroup\@empty \let\math@egroup\macc@set@skewchar
\mathsurround\z@ \frozen@everymath{\mathgroup\macc@group\relax}%
\macc@set@skewchar\relax
\let\mathaccentV\macc@nested@a
\if#31
\macc@nested@a\relax111{#1}%
\else
\def\gobble@till@marker##1\endmarker{}%
\futurelet\first@char\gobble@till@marker#1\endmarker
\ifcat\noexpand\first@char A\else
\def\first@char{}%
\fi
\macc@nested@a\relax111{\first@char}%
\fi
\endgroup
}
\makeatother

\newcommand\ee{\end{equation}}
\newcommand\be{\begin{equation}}
\newcommand\eea{\end{eqnarray}}
\newcommand\bea{\begin{eqnarray}}
\newcommand{\bsp}{\begin{split}}
\newcommand{\esp}{\end{split}}
\newcommand{\bit}{\begin{itemize}[leftmargin=*]}
\newcommand{\eit}{\end{itemize}}
\newcommand{\ben}{\begin{enumerate}[leftmargin=*]}
\newcommand{\een}{\end{enumerate}}

\renewcommand{\emph}{\textit}
\newcommand\eq[1]{Eq.~\eqref{eq:#1}}

\newcommand{\eqsII}[2]{Eqs.~\eqref{eq:#1}, \eqref{eq:#2}}
\newcommand{\eqsIII}[3]{Eqs.~\eqref{eq:#1}, \eqref{eq:#2}, \eqref{eq:#3}}
\newcommand{\eqsIV}[4]{Eqs.~\eqref{eq:#1}, \eqref{eq:#2}, \eqref{eq:#3}, \eqref{eq:#4}}

\newcommand{\iu}{\mathrm{i}}%\mkern1mu}
\newcommand{\eu}{\mathrm{e}}%\mkern1mu}
\newcommand{\dif}{\mathrm{d}}

\newcommand{\Tr}{\mathrm{Tr}\,}
\renewcommand{\vec}{\bm} 
\newcommand\vers[1]{\hat{\vec{#1}}}

\newcommand\eps{\varepsilon}

\def\<{\left\langle}
\def\>{\right\rangle}

\def\comment#1{}

\title{The EFT Likelihood for Large-Scale Structure in Redshift Space}

\author[a]{Giovanni Cabass}%,}%,}%\note{Also at Some University.}}

\affiliation[a]{Max-Planck-Institut f\"{u}r Astrophysik, %\\ 
Karl-Schwarzschild-Str. 1, 85741 Garching, Germany}

\emailAdd{gcabass@mpa-garching.mpg.de}

\abstract{\noindent We study the EFT likelihood for biased tracers in redshift space, 
for which the bias expansion of the galaxy velocity field $\vec{v}_g$ plays a fundamental role. 
The equivalence principle forbids stochastic contributions to $\vec{v}_g$ to survive at small $k$. 
Therefore, at leading order in derivatives the 
form of the likelihood ${\cal P}[\tilde{\delta}_g|\delta,\!\vec{v}]$ 
to observe a redshift-space galaxy overdensity $\tilde{\delta}_g(\tilde{\vec{x}})$ 
given a rest-frame matter and velocity fields 
$\delta(\vec{x})$, $\vec{v}(\vec{x})$ is fixed by the rest-frame noise. 
If this noise is Gaussian with constant power spectrum, 
${\cal P}[\tilde{\delta}_g|\delta,\!\vec{v}]$ is also a Gaussian in the 
difference between $\tilde{\delta}_g(\tilde{\vec{x}})$ and its 
bias expansion: redshift-space distortions only make the covariance depend on 
$\delta(\vec{x})$ and $\vec{v}(\vec{x})$. We then show how to match this result 
to perturbation theory, and that one can consistently neglect the field-dependent covariance 
if the bias expansion is stopped at second order in perturbations. We discuss qualitatively how 
this affects numerical implementations of the EFT-based forward modeling, and 
how the picture changes when the survey window function is taken into account.}

\begin{document}
\maketitle
\flushbottom

%\cleardoublepage

%-------------------------------------------------------------------------------------------------------

%********************************************************
\section{Introduction and summary of main results}
\label{sec:introduction}
%********************************************************

\noindent The effective field theory (EFT) of large-scale structure (LSS), 
EFTofLSS hereafter, allows for a controlled incorporation of the effects of fully nonlinear 
structure formation on small scales in the framework of cosmological perturbation 
theory \cite{Baumann:2010tm,Carrasco:2012cv}. This is especially important when 
attempting to infer cosmological information from observed biased tracers 
such as galaxies, quasars, galaxy clusters, the Lyman-$\alpha$ forest, and 
others (see \cite{Desjacques:2016bnm} for a review; in the following we 
will always refer to the tracers as ``galaxies'' for simplicity). 

The prediction for the galaxy density field $\delta_g(\vec{x},\tau) = n_g(\vec{x},\tau)/\widebar{n}_g(\tau)-1$ 
can be broken into two parts: a ``deterministic'' part $\delta_{g,{\rm det}}$ which captures the 
modulation of the galaxy density by long-wavelength perturbations, and a stochastic 
residual which fluctuates due to the small-scale initial conditions. When integrating 
out small-scale modes, this effectively leads to a noise in the galaxy density. 

Recently, Refs.~\cite{Schmidt:2018bkr,Cabass:2019lqx} presented a derivation of the likelihood of the 
entire galaxy density field $\delta_g(\vec{x},\tau)$ given the nonlinear matter density 
field (i.e.~the matter field evolved via gravity), in the context of the EFT. This result 
offers several advantages over approaches that aim at constraining a finite number of correlation functions: 
\begin{itemize}[leftmargin=*] 
\item It puts the stochasticity of galaxies and the deterministic bias expansion on the same footing, 
showing that the form of the conditional likelihood is determined by the properties of the noise 
(such as the fact that, in first approximation, it is Gaussian with constant power spectrum 
on scales where the EFTofLSS can be applied). 
\item The likelihood is given in terms of the fully nonlinear density field, which can be predicted for example using 
N-body simulations, and thus isolates the truly uncertain aspects of the observed galaxy density. 
\item The likelihood is given by the functional Fourier transform of the generating functional. 
Since the latter contains all correlation functions, the derivation of \cite{Cabass:2019lqx} provides a 
correspondence between different terms in the likelihood and correlation functions. 
\item The conditional likelihood of the galaxy density field given the evolved matter density 
field is precisely the key ingredient required in full Bayesian (``forward-modeling'') inference approaches 
\cite{1995MNRAS.272..885F,2010MNRAS.406...60J,2010MNRAS.409..355J,2013MNRAS.432..894J,Wang:2014hia,2015MNRAS.446.4250A}, 
and can be employed there directly \cite{Schmidt:2018bkr,Elsner:2019rql,2020arXiv200406707S} 
(see \cite{1989ApJ...336L...5B,Schmittfull:2017uhh,Seljak:2017rmr,Modi:2019hnu} for related approaches). 
\end{itemize} 

So far the EFT likelihood has been used with rest-frame halo catalogs only \cite{Elsner:2019rql,2020arXiv200406707S}. 
However, we need to account for the mapping from rest-frame quantities to observations (for which 
we loosely use the term ``projection effects'') if we want to apply the EFT-based Bayesian forward modeling to real data. 

Let us consider what happens on sub-horizon scales (that will be the focus of this work). 
There, projection effects reduce to redshift-space distortions (RSDs), 
so what is necessary is an expression for the likelihood in redshift space. Luckily, 
the EFTofLSS in redshift space has been thoroughly 
investigated already, see e.g.~\cite{Lewandowski:2015ziq,Perko:2016puo,Ding:2017gad,delaBella:2018fdb,Desjacques:2018pfv}. 
The main ingredient we need to move from the rest-frame coordinates to redshift space is the galaxy 
velocity field $\vec{v}_g$, and the equivalence principle strongly constrains the form of the 
EFT counterterms. Consider for example the well-known Kaiser formula for the redshift-space power spectrum 
\cite{Kaiser:1987qv}: the fact that we can use its quadrupole to test the growth rate without 
complications of bias is because on large scales the deterministic part of the galaxy velocity follows the 
matter velocity, $\smash{\vec{v}_{g,{\rm det}} = \vec{v}}$. The first EFT corrections enter at the derivative 
level, via operators like $\smash{\vec{v}_{g,{\rm det}}\supset\beta_{\nabla^2\vec{v}}\nabla^2\vec{v}}$. 

Something similar happens for the stochasticity in the galaxy velocity, $\vec{\varepsilon}_{v}$ 
in the notation of \cite{Desjacques:2016bnm}. The equivalence principle forbids its power spectrum 
to have a constant part in the large-scale limit. For our purposes, the most important consequence 
is that the form of the conditional likelihood is fixed fully by the properties of the 
noise for the galaxy density in the rest frame, at leading order in a derivative expansion: 
no additional stochasticity is involved when we move to redshift space. 

The main goal of the paper is to use this fact to compute the conditional EFT likelihood 
$\smash{{\cal P}[\tilde{\delta}_g|\delta,\!\vec{v}]}$ to observe a redshift-space galaxy 
overdensity $\smash{\tilde{\delta}_g(\tilde{\vec{x}})}$ given a rest-frame matter and 
velocity fields $\delta(\vec{x})$, $\vec{v}(\vec{x})$ at all orders in the deterministic 
bias expansion for $\smash{\tilde{\delta}_g}$. Then, we study how our result connects 
to the perturbative treatment of the EFTofLSS, and discuss the difficulties (and ways 
around them) of implementing a cutoff on the galaxy and matter fields (central to the 
EFT-based forward modeling) while retaining the full, nonperturbative form of the conditional 
likelihood. Finally, we look into what are the corrections coming from the noise in the galaxy velocity field 
and briefly touch on the survey window function. The short section below contains a more detailed summary.

%********************************************************
\subsection*{Outline and summary of main results}
\label{subsec:outline_and_summary}
%********************************************************

\noindent The outline of the paper, and a summary of the main results, is as follows. 
\begin{description}[leftmargin=*]
\item[Section~\ref{sec:notation_and_review}] contains a quick review of RSDs following Section~9.3 of 
\cite{Desjacques:2016bnm}. Then we summarize our notation and conventions and review the derivation 
of the rest-frame likelihood following \cite{Schmidt:2018bkr,Cabass:2019lqx}. 
\item[Section~\ref{sec:main_result}] derives the main result of the paper. 
We obtain the conditional likelihood for the redshift-space galaxy density at leading 
order in a derivative expansion but at all orders in perturbations 
(Sections~\ref{subsec:coordinate_change} and \ref{subsec:integrating_out_noise}). The result is 
\begin{equation}
\label{eq:intro}
{\cal P}[\tilde{\delta}_g|\delta,\!\vec{v}] = \Bigg(\prod_{\tilde{\vec{x}}}\sqrt{ 
\frac{\tilde{J}[\delta,\!\vec{v}](\tilde{\vec{x}})}{2\pi P_{\eps_g}^{\{0\}}}}\,\Bigg)\exp\Bigg( 
{-\frac{1}{2}}\int\dif^3\tilde{x}\,\frac{\big( 
\tilde{\delta}_g(\tilde{\vec{x}}) - \tilde{\delta}_{g,{\rm det}}[\delta,\!\vec{v}](\tilde{\vec{x}})\big)^2} 
{P_{\eps_g}^{\{0\}}/\tilde{J}[\delta,\!\vec{v}](\tilde{\vec{x}})}\Bigg)\,\,, 
\end{equation}
where: 
\begin{itemize}[leftmargin=*]
\item following the notation of \cite{Desjacques:2016bnm} we denote the redshift-space coordinates by $\tilde{\vec{x}}$; 
\item $\smash{\tilde{\delta}_g(\tilde{\vec{x}})}$ are the data in redshift space; 
\item $\smash{P_{\eps_g}^{\{0\}}}$ is the noise power spectrum, usually parameterized 
as $\alpha/\widebar{n}_g$ for a dimensionless $\alpha$; 
\item $\smash{\tilde{\delta}_{g,{\rm det}}}$ is the deterministic bias expansion for the redshift-space galaxy overdensity. 
It is obtained by transforming to redshift space its rest-frame analog $\smash{\delta_{g,{\rm det}}}$. 
For this reason it depends both on $\delta$ and on the matter velocity $\vec{v}$; 
\item finally, we denote by $\smash{\tilde{J}}$ the Jacobian of the transformation from the rest frame to redshift space. 
In the distant-observer approximation $\smash{\vers{n} = {\rm const.}}$ ($\smash{\vers{n}}$ being the line of sight) it is equal to 
$\smash{1+\vers{n}\cdot\partial_\parallel\vec{v}_g(\vec{x})/{\cal H}}$ evaluated at $\vec{x}=\vec{x}(\tilde{\vec{x}})$, 
and with $\smash{\vec{v}_g = \vec{v}_{g,{\rm det}}[\delta,\!\vec{v}]}$. It depends on the matter overdensity and 
the matter velocity field because the deterministic bias expansion for the galaxy velocity and the coordinate change 
$\smash{\vec{x}=\vec{x}(\tilde{\vec{x}})}$ depend on them. Writing explicitly the Jacobian in the distant-observer 
approximation, \eq{intro} then becomes 
\begin{equation}
\label{eq:intro_distant_observer} 
\begin{split}
{\cal P}[\tilde{\delta}_g|\delta,\!\vec{v}] &= \Bigg(\prod_{\tilde{\vec{x}}}\sqrt{ 
\frac{1+\vers{n}\cdot\partial_\parallel\vec{v}_{g,{\rm det}}[\delta,\!\vec{v}](\vec{x}(\tilde{\vec{x}})) 
/ {\cal H}}{2\pi P_{\eps_g}^{\{0\}}}}\,\Bigg) \\
&\;\;\;\;\times\exp\Bigg( 
{-\frac{1}{2}}\int\dif^3\tilde{x}\,\frac{\big( 
\tilde{\delta}_g(\tilde{\vec{x}}) - \tilde{\delta}_{g,{\rm det}}[\delta,\!\vec{v}](\tilde{\vec{x}})\big)^2} 
{P_{\eps_g}^{\{0\}}}\Bigg\{1+\frac{\vers{n}\cdot\partial_\parallel\vec{v}_{g,{\rm det}}[\delta,\!\vec{v}](\vec{x}(\tilde{\vec{x}})) 
}{{\cal H}}\Bigg\}\Bigg)\,\,. 
\end{split}
\end{equation} 
\end{itemize} 
Section~\ref{subsec:zero_noise_normalization_etc} contains multiple checks of \eq{intro}, such as the fact that 
we must obtain a Dirac delta functional setting $\smash{\tilde{\delta}_g}$ equal to $\smash{\tilde{\delta}_{g,{\rm det}}}$ 
when the noise goes to zero (which, given that the likelihood is Gaussian in the data $\smash{\tilde{\delta}_g}$, 
is equivalent to requiring that its integral over $\smash{\tilde{\delta}_g}$ equals $1$). This is explicit in 
\eq{intro}, thanks to the fact that the overall factor in front of the exponential is written as a product over redshift-space 
coordinates $\smash{\tilde{\vec{x}}}$. Finally, we show that it is possible to augment \eq{intro} to include 
the modulation of the rest-frame noise by long-wavelength operators (effectively equivalent to a 
stochasticity of the bias coefficients in $\delta_{g,{\rm det}}$). 
This amounts to replacing $\smash{P_{\eps_g}^{\{0\}}}$ with a nonnegative ``field-dependent covariance''. 
\item[Section~\ref{sec:cutoffs_and_perturbativity}.] 
The conditional likelihood of \eq{intro} above is written in real space. 
This is necessary if we want an expression that is valid at all orders in perturbations, 
since stopping at leading order in derivatives means that we always deal with local functionals. 
This can, however, obscure the link with perturbative approaches. In Sections~\ref{subsec:cutoffs} 
and \ref{subsec:perturbativity} we show that once we cutoff the Fourier modes of the fields at a 
scale $\Lambda$, it is straightforward to recover the expansion parameters of the EFTofLSS. This 
procedure allows us also to address the positivity of the Jacobian $\smash{\tilde{J}}$ in \eq{intro}. 
The use of filtered fields (both galaxy and matter ones) is central to current applications of 
the EFT likelihood \cite{Elsner:2019rql,2020arXiv200406707S}: in Section~\ref{subsec:covariance} 
we discuss how the presence of the field-dependent Jacobian makes it complicated to retain a closed form 
of the likelihood and have it normalized with respect to the data, i.e.~the galaxy field, once this is cut 
at $\Lambda$. Fortunately we also show that, as long as we stop at second order in the bias expansion, it 
is consistent to neglect the dependence of $\smash{\tilde{J}}$ on $\smash{\delta}$ and $\smash{\vec{v}}$, 
hence allowing an analytical normalization of the likelihood. We also briefly investigate how 
it is possible to bypass this problem while keeping the full dependence of the Jacobian on the 
matter density and velocity fields. 
\item[Section~\ref{sec:velocity_noise}] studies the impact of the stochasticity in the galaxy velocity field, 
$\smash{\vec{v}_g = \vec{v}_{g,{\rm det}} + \vec{\eps}_v}$. As discussed in the introduction, this is expected 
to be subleading on large scales. After reviewing the power spectrum of $\smash{\vec{\eps}_v}$, we confirm this 
via explicit computation in a way similar to Section~\ref{subsec:perturbativity}. Fig.~\ref{fig:scalings} 
summarizes the relative importance of these contributions with respect to the ones studied in 
Sections~\ref{subsec:cutoffs} and \ref{subsec:perturbativity}. 
\item[Section~\ref{sec:conclusions}] concludes the paper with a brief discussion of astrophysical selection 
effects and, especially, of the survey window function. 
\item[Appendices~\ref{app:integrating_out_noise_appendix}, \ref{app:functional_change_of_coordinates} 
and \ref{app:velocity_noise}] contain some details on the calculations of Sections~\ref{sec:main_result}, 
\ref{sec:cutoffs_and_perturbativity} and~\ref{sec:velocity_noise}, respectively. 
\end{description}

%********************************************************
\section{RSDs, notation and review of the rest-frame likelihood}
\label{sec:notation_and_review}
%********************************************************

%********************************************************
\subsection{Redshift-space distortions}
\label{subsec:rsds}
%********************************************************

\noindent We start by a quick review of redshift-space distortions (we refer e.g.~to 
\cite{Bernardeau:2001qr,Shun_on_RSDs,Vlah:2018ygt} and references therein for additional 
details). In this section we follow closely the notation of \cite{Desjacques:2016bnm} 
(see their Section~9.3), while from the next section to the end of the paper we will 
use a modified version that better highlights how the transformation 
to redshift space depends on the matter density and velocity fields. 

All tracers of large-scale structure are effectively observed via photon arrival directions ($\vers{n}$) and redshifts ($z$), 
inferred from the shift in frequency of the observed spectral energy distribution of the galaxy relative to 
the rest-frame frequency. Hence, an essential ingredient in the interpretation of large-scale structure is the mapping 
from rest-frame quantities to observations (see \cite{Jeong:2014ufa} for a concise recent review of the subject). 

We can relate the observed position $\smash{(z,\vers{n})}$ to the position of the galaxy 
in a global coordinate system $\smash{x^\mu = (\eta,\vec{x})}$ by solving the geodesic 
equation from the observer's location to the source, given the photon momentum at the 
observer specified by $\smash{(z,\vers{n})}$. We can also associate a ``fiducial'' 
position $\smash{\tilde{x}^\mu = (\tilde{\eta},\tilde{\vec{x}})}$ to the 
galaxy by solving the same geodesic equation in a purely FLRW spacetime.\footnote{When 
also fixing a fiducial cosmology we obtain the additional Alcock-Paczy{\'n}ski distortions 
\cite{Alcock:1979mp}.} The difference $\smash{\Delta x}$ between $x$ and $\smash{\tilde{x}}$ effectively 
defines a coordinate transformation from observed to true galaxy positions, and we can find the expression for 
the observed galaxy density by computing how the zeroth component of the galaxy number current transforms under it. 

On subhorizon scales $k\gg{\cal H}$, which will be main focus of this work, 
the gravitational redshift terms (which involve the Newtonian potential directly) are negligible, as is 
the component of $\Delta\vec{x}$ orthogonal to the line of sight. This implies that the coordinate 
shift from the rest frame to redshift space is purely spatial and parallel to the line of sight. It is given by 
\begin{equation}
\label{eq:RS_review-1}
\tilde{\vec{x}} = \vec{x} + u_{\parallel}(\vec{x})\,\vers{n}(\vec{x})\,\,, 
\end{equation}
where 
\begin{equation}
\label{eq:RS_review-2}
\vec{u}(\vec{x}) = \frac{\vec{v}_g(\vec{x})}{{\cal H}}\,\,, 
\quad u_{\parallel}(\vec{x}) = \vers{n}(\vec{x})\cdot\vec{u}(\vec{x})\,\,,\quad\vers{n}(\vec{x})=\frac{\vec{x}}{\abs{\vec{x}}}\,\,. 
\end{equation} 
Here $\vec{v}_g$ is the galaxy velocity, and for simplicity of notation we will not add a subscript to $u_{\parallel}$. 

If we make the assumption that the transformation of \eq{RS_review-1} is one-to-one, which is true on perturbative scales 
(we will come back to this in Section~\ref{subsec:perturbativity}), 
we can obtain a fully nonlinear expression for the observed galaxy density perturbation as 
\begin{equation}
\label{eq:RS_review-3}
\tilde{\delta}_g(\tilde{\vec{x}}) = \frac{1+\delta_g(\vec{x})}{1+\partial_\parallel u_\parallel(\vec{x})} - 1\,\,. 
\end{equation}
On the right-hand side we intend everything evaluated at $\smash{\vec{x}=\vec{x}(\tilde{\vec{x}})}$, and we have used the relation 
\begin{equation}
\label{eq:RS_review-4}
\abs[\bigg]{\frac{\partial x^i}{\partial\tilde{x}^j}} = \abs[\bigg]{\delta^j_i 
+ \hat{n}^j\frac{\partial u_\parallel(\vec{x})}{\partial x^i}}^{-1}
= \frac{1}{1+\partial_\parallel u_\parallel(\vec{x})}\,\,, 
\end{equation} 
where again on the right-hand side everything is evaluated at $\smash{\vec{x}=\vec{x}(\tilde{\vec{x}})}$, 
and $\smash{\partial_\parallel}$ is defined as $\smash{\vers{n}\cdot\vec{\nabla}}$ in \eqsII{RS_review-3}{RS_review-4}. 
This relation holds in the distant-observer approximation, where we approximate $\smash{\vers{n}}$ as slowly-varying 
over the survey volume. The final result of this paper, \eq{intro}, is not dependent on this approximation: as we will see in 
the next sections it will hold whatever the form of the Jacobian $\smash{\abs{\partial x^i/\partial\tilde{x}^j}}$ is.

%********************************************************
\subsection{Notation and conventions}
\label{subsec:notation_and_conventions}
%********************************************************

\noindent For the purposes of this paper it is fundamental to 
keep in mind the fact that the transformation to redshift space 
depends on galaxy velocity field. Let us then write \eq{RS_review-1} as 
\begin{equation}
\label{eq:notation-1}
\vec{x} = \mathscr{R}[\vec{v}_g](\tilde{\vec{x}})\,\,,\quad\tilde{\vec{x}} = \mathscr{R}^{-1}[\vec{v}_g](\vec{x})\,\,. 
\end{equation}
The Jacobian of the transformation from redshift space to the rest-frame coordinates is given by 
\begin{equation}
\label{eq:notation-2}
\abs[\bigg]{\frac{\partial \mathscr{R}^{-1}[\vec{v}_g](\vec{x})}{\partial\vec{x}}} = 
1 + \partial_\parallel u_\parallel(\vec{x}) \equiv J[\vec{v}_g](\vec{x})\,\,,
\end{equation}
while the Jacobian of the inverse transformation is given by 
\begin{equation}
\label{eq:notation-3-A}
\text{$\abs[\bigg]{\frac{\partial \mathscr{R}[\vec{v}_g](\tilde{\vec{x}})}{\partial\tilde{\vec{x}}}} = 
\frac{1}{\tilde{J}[\vec{v}_g](\tilde{\vec{x}})}\,\,,\quad$ with 
$\quad\tilde{J}[\vec{v}_g](\tilde{\vec{x}}) = J[\vec{v}_g]\big(\mathscr{R}[\vec{v}_g](\tilde{\vec{x}})\big)\,\,.$} 
\end{equation} 
Consequently, \eq{RS_review-3} becomes 
\begin{equation}
\label{eq:notation-3-B}
\tilde{\delta}_g(\tilde{\vec{x}}) = \frac{1+\delta_g\big(\mathscr{R}[\vec{v}_g](\tilde{\vec{x}})\big)}
{\tilde{J}[\vec{v}_g](\tilde{\vec{x}})} - 1\,\,. 
\end{equation}

Let us now move to the bias expansion, that will allow us to streamline the notation of \eq{notation-3-B} a bit. 
First, we recap the bias expansion for the galaxy density field in the rest frame. 
If we define the nonlinear matter field as $\delta$, we can write the deterministic bias relation as 
\begin{equation}
\label{eq:notation-4}
\delta_g(\vec{x}) = \delta_{g,{\rm det}}[\delta](\vec{x})\,\,,
\end{equation}
where the functional $\delta_{g,{\rm det}}[\delta]$ contains all the 
operators constructed from the nonlinear matter field. 
Let us write it as 
\begin{equation} 
\label{eq:notation-5}
\delta_{g,{\rm det}}[\delta] = \sum_{O}b_{O}\,{O}[\delta]\,\,. 
\end{equation} 
Here we use the basis of \cite{Mirbabayi:2014zca} (see also Sections~2.2--2.5 of \cite{Desjacques:2016bnm}, 
and see \cite{Senatore:2014eva} for an alternative basis) to write the bias expansion at a fixed time. 
Then, in the rest frame and up to second order in perturbations (and leading order in derivatives) we have 
\begin{equation}
\label{eq:notation-6}
\delta_{g,{\rm det}}[\delta](\vec{x}) = b_1\delta(\vec{x}) + \frac{b_2}{2}\delta^2(\vec{x}) + b_{K^2}K^2[\delta]\,\,, 
\end{equation} 
where $K^2 = K_{ij}K^{ij}$ and the tidal field $K_{ij}[\delta]$ is equal to $(\partial_i\partial_j/\nabla^2 - \delta_{ij}/3)\delta$. 

We can then look at the galaxy velocity $\vec{v}_g$. The matter velocity field $\vec{v}$ 
plays a fundamental role in its deterministic bias expansion. 
More precisely, the equivalence principle ensures that at leading order in derivatives we have 
\begin{equation}
\label{eq:notation-7}
\vec{v}_{g,{\rm det}}[\delta,\!\vec{v}](\vec{x}) = \vec{v}(\vec{x})\,\,, 
\end{equation} 
i.e.~we have $\beta_v = 1$ (we use the same notation as \cite{Desjacques:2016bnm} for velocity-bias parameters). 
Terms schematically the form $(\delta\cdots\delta)\vec{v}$, or generically 
$O[\delta]\vec{v}$, are likewise forbidden. The leading correction takes the form \cite{Desjacques:2010gz,Baldauf:2014fza} 
\begin{equation}
\label{eq:notation-8}
\vec{v}_{g,{\rm det}}[\delta,\!\vec{v}](\vec{x})\supset \beta_{\nabla^2\vec{v}}\nabla^2\vec{v}(\vec{x})\,\,, 
\end{equation}
which is degenerate with $\vec{\nabla}\delta$ at linear order in perturbations. 
Notice that on the right-hand side we have allowed for a generic functional dependence on both the 
nonlinear matter field $\delta$ and the nonlinear velocity field $\vec{v}$: a complete enumeration 
of all the operators contained in $\vec{v}_{g,{\rm det}}[\delta,\!\vec{v}]$ at leading order in derivatives, 
together with the proof that also for the velocity field it is possible to write a bias expansion at a fixed time, 
can be found in \cite{Mirbabayi:2014zca} (see also Appendix~B.5 of \cite{Desjacques:2016bnm}). 

As the reader might have noticed, in this section (and in Section~\ref{subsec:rsds} as well) 
we have focused on the deterministic bias expansion for the galaxy density and velocity fields. 
A more detailed discussion of the noise $\vec{v}_g-\vec{v}_{g,{\rm det}}[\delta,\!\vec{v}]$ 
is left to Section~\ref{sec:velocity_noise}: for the moment we emphasize that, in a derivative expansion, the leading source of noise is 
only the rest-frame noise $\eps_g$ for $\delta_g$, whose power spectrum ($\sim k^0$ on large scales) is usually parameterized 
as $\alpha/\widebar{n}_g$ for some dimensionless $\alpha$ expected to be of order $1$. This is also reviewed in the next section. 

The fact that we focus on the deterministic bias expansion for $\vec{v}_g$ 
allows us to simplify the notation greatly. Until Section~\ref{sec:velocity_noise} 
we will always understand $\smash{\vec{v}_g=\vec{v}_{g,{\rm det}}[\delta,\!\vec{v}]}$ 
in the coordinate transformation $\mathscr{R}$: therefore we will drop the functional 
dependence of the latter on $\vec{v}_g$, i.e.~ 
\begin{equation} 
\label{eq:notation-9-A} 
\mathscr{R}\big[\vec{v}_{g,{\rm det}}[\delta,\!\vec{v}]\big]\equiv\mathscr{R}\,\,. 
\end{equation}
Similarly, since the Jacobian of the coordinate change also depends on $\smash{\delta}$ 
and $\smash{\vec{v}}$ via $\smash{\vec{v}_{g,{\rm det}}}$, we write 
\begin{equation}
\label{eq:notation-9-B}
J\big[\vec{v}_{g,{\rm det}}[\delta,\!\vec{v}]\big](\vec{x})\equiv J[\delta,\!\vec{v}](\vec{x})\,\,, 
\end{equation} 
and 
\begin{equation}
\label{eq:notation-9-C}
\tilde{J}\big[\vec{v}_{g,{\rm det}}[\delta,\!\vec{v}]\big](\tilde{\vec{x}}) 
= 
J\big[\vec{v}_{g,{\rm det}}[\delta,\!\vec{v}]\big]\big(\mathscr{R}(\tilde{\vec{x}})\big) 
\equiv 
\tilde{J}[\delta,\!\vec{v}](\tilde{\vec{x}})\,\,,
\end{equation}
which is consistent with the notation in \eq{intro}. Finally, we also define 
\begin{equation}
\label{eq:notation-10}
\tilde{\delta}_{g,{\rm det}}[\delta,\!\vec{v}](\tilde{\vec{x}}) = 
\frac{1+\delta_{g,{\rm det}}[\delta]\big(\mathscr{R}(\tilde{\vec{x}})\big)}
{\tilde{J}[\delta,\!\vec{v}](\tilde{\vec{x}})} - 1\,\,. 
\end{equation} 
This functional is what appears in \eq{intro}. 

We conclude this section with a recap of our notation for functional derivatives (importantly, 
both $\vec{x}$ and $\tilde{\vec{x}}$ are cartesian coordinates, so the following relations work 
equally well in the rest-frame coordinates and redshift space). Given a field $\chi(\vec{x})$, we have 
\begin{equation}
\label{eq:notation-11}
\frac{\partial\chi(\vec{x})}{\partial\chi(\vec{y})} = \delta^{(3)}_{\rm D}(\vec{x}-\vec{y})\,\,. 
\end{equation} 
This is the generalization of $\partial x^i/\partial x^j = \delta^i_j$. 
The right-hand side is dimensionful since we need to satisfy the relation 
\begin{equation}
\label{eq:notation-12}
\frac{\partial}{\partial\chi(\vec{y})}\int\dif^3x\,\chi(\vec{x}) = 1\,\,,
\end{equation}
i.e.~the equivalent of $\partial\big(\sum_i x^i\big)/\partial x^j = 1$. We define the functional Dirac delta by 
\begin{equation}
\label{eq:notation-13}
\int{\cal D}\chi\,F[\chi]\,\delta^{(\infty)}_{\rm D}(\chi-\varphi) = F[\varphi] 
\end{equation} 
for any functional $\smash{F[\chi]}$. Given the functional measure $\smash{{\cal D}\chi = \prod_{\vec{x}}{\rm d}\chi(\vec{x})}$, 
we can see that for practical purposes $\smash{\delta^{(\infty)}_{\rm D}(\chi-\varphi)}$ is a product of one-dimensional 
Dirac delta functions of $\smash{\chi(\vec{x})-\varphi(\vec{x})}$ at each $\smash{\vec{x}}$. 
Analogous definitions (with $\smash{\delta^{(3)}_{\rm D}(\vec{x}-\vec{x}')\to 
(2\pi)^3\delta^{(3)}_{\rm D}(\vec{k}+\vec{k}')}$, etc.) hold for functionals of momentum-space fields.

%********************************************************
\subsection{Review of the rest-frame EFT likelihood}
\label{subsec:eft_likelihood_review}
%********************************************************

\noindent In this section we review the results of \cite{Cabass:2019lqx}, where the rest-frame EFT likelihood was derived 
under the assumption of Gaussian noise and no stochasticity of the bias coefficients. 

The difference between $\delta_g$ and $\delta_{g,{\rm det}}[\delta]$ that arises from integrating out short-scale modes 
that cannot be described within the EFT is captured by a noise $\eps_g(\vec{x})$. Let us assume that the 
noise is Gaussian with power spectrum $P_{\eps_g}(k)$. Locality (i.e.~the fact that the error we make in 
describing galaxy clustering via \eq{notation-5} at $\vec{x}_1$ and $\vec{x}_2$ 
is uncorrelated in the limit of large $\smash{\abs{\vec{x}_1-\vec{x}_2}}$) 
and the absence of preferred directions impose that the noise power spectrum 
is analytic in $\smash{k^2 = \abs{\vec{k}}^2}$, i.e.~ 
\begin{equation}
\label{eq:like_review-1}
P_{\eps_g}(k) = P^{\{0\}}_{\eps_g} + P^{\{2\}}_{\eps_g}k^2 + \dots\,\,.
\end{equation}
The coefficients $\smash{P^{\{n\}}_{\eps_g}}$ have dimensions of a length to the power $n+3$: $\smash{P^{\{0\}}_{\eps_g}}$ 
fixes the size of the noise (this is what is typically taken to be $\alpha/\bar{n}_g$), while we expect that for $n\geq 2$ we have 
\begin{equation}
\label{eq:like_review-2}
\frac{P^{\{n\}}_{\eps_g}}{P^{\{0\}}_{\eps_g}}\sim R_\ast^n\,\,, 
\end{equation}
where $\smash{R_\ast}$ is the typical nonlocality scale of galaxy formation. For dark matter halos, $\smash{R_\ast}$ is expected to be 
of order of the halo Lagrangian radius $\smash{R(M_h)}$ or of order of the nonlocality scale for matter 
$\smash{\sim 1/k_{\rm NL}}$ (that is the scale at which the dimensionless linear matter power spectrum 
becomes of order one), whichever is larger (see e.g.~\cite{Smith:2006ne,Hamaus:2010im,Baldauf:2013hka,Chan:2014qka} 
for a discussion of deviations of $\smash{\alpha}$ from $\smash{1}$ and for the scale dependence of $\smash{P_{\eps_g}(k)}$). 

Let us then take a wavenumber $\Lambda$ smaller than $1/R_\ast$. We can split 
the noise field in a short-wavelength part and a long-wavelength part. More 
precisely, the short-wavelength one is obtained by subtracting 
\begin{equation}
\label{eq:like_review-3}
\eps_{g,\Lambda}(\vec{k}) = \eps_g(\vec{k})\,\Theta_{\rm H}\big(\Lambda^2-k^2\big) 
\end{equation}
from $\eps_{g}(\vec{k})$, where $\smash{\Theta_{\rm H}}$ is the Heaviside theta function. 
Since we are assuming the noise to be Gaussian the likelihood for the 
short modes and the long modes factorizes, as does the functional measure ${\cal D}\eps_g$. 
Given that we cannot reliably describe short-wavelength modes, we can just integrate out the short-wavelength 
component of the noise, and remain with a likelihood for $\eps_{g,\Lambda}(\vec{k})$ only. 

What is this likelihood? Since we have chosen $\Lambda$ such that the higher-derivative 
terms of \eq{like_review-1} are negligible, we can write it as 
\begin{equation}
\label{eq:like_review-4}
{\cal P}[\eps_g] = \Bigg(\prod_{\abs{\vec{k}}\leq\Lambda}\sqrt{\frac{1}{2\pi P^{\{0\}}_{\eps_g}}}\,\Bigg) 
\exp\Bigg({-\frac{1}{2}}\int_{\vec{k}}\frac{\abs{\eps_{g,\Lambda}(\vec{k})}^2}{P^{\{0\}}_{\eps_g}}\Bigg)\,\,. 
\end{equation} 
The normalization of \eq{like_review-4} is such that, if $\smash{P^{\{0\}}_{\eps_g}\to 0}$, we recover a Dirac delta 
functional that sets $\eps_{g,\Lambda}$ to zero. Before proceeding, let us discuss the 
assumption of Gaussian likelihood (we go back to this in Section~\ref{subsec:perturbativity}, 
see also Fig.~\ref{fig:scalings}). Is there a small parameter that allows us to expand around 
a Gaussian? The fact that we are restricting ourselves to long wavelengths ensures that 
higher-order $n$-point functions of the noise are suppressed. For example, consider the noise bispectrum. 
Via the same arguments that lead to \eq{like_review-1}, we can take it to be 
a constant on large scales. By dimensional analysis we can take this constant to 
be a dimensionless $S_{\epsilon_g}$ times $\smash{P^{\{0\}}_{\eps_g}}$ squared ($S_{\epsilon_g}$ is $1$ 
for a Poisson likelihood, for example). The non-Gaussianity of the noise likelihood, then, 
is controlled by $S_{\epsilon_g}$ times the typical size of a noise fluctuation on a scale $\Lambda$, 
which scales as 
\begin{equation}
\label{eq:noise_scaling}
\sqrt{P^{\{0\}}_{\eps_g}\Lambda^3}\,\,. 
\end{equation} 
Hence, as long as $\smash{\Lambda}$ is smaller than 
$\smash{S_{\epsilon_g}^{-2/3}(\bar{n}_g/\alpha)^{1/3}}$, we are justified in expanding around a Gaussian likelihood. 

Let us then multiply this likelihood by a (``Fourier-space'') Dirac delta functional 
\begin{equation}
\label{eq:like_review-5}
\delta^{(\infty)}_{\rm D}\big(\delta_{g,\Lambda}(\vec{k}) - \delta_{g,{\rm det},\Lambda}[\delta_\Lambda](\vec{k}) 
- \eps_{g,\Lambda}(\vec{k})\big)\,\,. 
\end{equation} 
Here we have cut both $\smash{\delta_g}$ and $\smash{\delta_{g,{\rm det}}}$ at $\Lambda$, 
and we have constructed the deterministic galaxy field from the matter field cut also at $\Lambda$. 
This is the same procedure that was originally described in \cite{Schmidt:2018bkr}. 
We will come back to it in Section~\ref{subsec:cutoffs}. 

If we now functionally integrate over $\eps_{g,\Lambda}$, we obtain the conditional likelihood for 
the galaxy field given the matter field, i.e.~ 
\begin{equation}
\label{eq:like_review-6}
{\cal P}[\delta_{g,\Lambda}|\delta_\Lambda] = 
\Bigg(\prod_{\abs{\vec{k}}\leq\Lambda}\sqrt{\frac{1}{2\pi P^{\{0\}}_{\eps_g}}}\,\Bigg) 
\exp\Bigg({-\frac{1}{2}}\int_{\abs{\vec{k}}\leq \Lambda} 
\frac{\abs{\delta_g(\vec{k}) - \delta_{g,{\rm det}}[\delta_\Lambda](\vec{k})}^2}{P^{\{0\}}_{\eps_g}}\Bigg)\,\,. 
\end{equation}
Here we have used the fact that the data and the deterministic galaxy density field have both support for $\abs{\vec{k}}\leq\Lambda$ 
to remove the cutoff from the fields themselves and replace it by a cutoff in the integral $\int_{\vec{k}}$, using the fact that 
these two fields appear quadratically in the likelihood. 

Thanks to the fact that both the galaxy field and its deterministic expression in terms of $\smash{\delta_\Lambda}$ 
appear quadratically in the exponent of \eq{like_review-6}, we can switch to real space. More precisely, we can write 
\begin{equation}
\label{eq:like_review-7}
{\cal P}[\delta_{g,\Lambda}|\delta_\Lambda] = \Bigg(\prod_{\vec{x}}\sqrt{\frac{1}{2\pi P_{\eps_g}^{\{0\}}}}\,\Bigg)\exp\Bigg(
{-\frac{1}{2}}\int\dif^3x\,\frac{\big(\delta_{g,\Lambda}(\vec{x}) - 
\delta_{g,{\rm det},\Lambda}[\delta_\Lambda](\vec{x})\big)^2}{P_{\eps_g}^{\{0\}}}\Bigg)\,\,, 
\end{equation}
where the ``$\Lambda$'' subscripts stand for the fact that: 
(i) we cut the field $\delta_g$ in Fourier space and transform it back to real space; 
(ii) we construct $\delta_{g,{\rm det}}$ from $\delta_\Lambda$, we cut it in Fourier space, and then transform it to real space. 
We then take the difference between $\smash{\delta_{g,\Lambda}(\vec{x})}$ and 
$\smash{\delta_{g,{\rm det},\Lambda}[\delta_\Lambda](\vec{x})}$, square it, and integrate it over all $\vec{x}$. 
Effectively, this tells us that it makes sense to write 
\begin{equation}
\label{eq:like_review-8}
{\cal P}[\delta_g|\delta] = \Bigg(\prod_{\vec{x}}\sqrt{\frac{1}{2\pi P_{\eps_g}^{\{0\}}}}\,\Bigg)\exp\Bigg(
{-\frac{1}{2}}\int\dif^3x\,\frac{\big(\delta_g(\vec{x})-\delta_{g,{\rm det}}[\delta](\vec{x})\big)^2}{P_{\eps_g}^{\{0\}}}\Bigg)\,\,, 
\end{equation}
if we assume that the fields $\smash{\delta}$ and $\smash{\delta_{g,{\rm det}}[\delta]}$ 
appearing in the integral above are cut at a scale longer than $1/R_\ast$, defined by \eq{like_review-2}. 
In \eqsII{like_review-7}{like_review-8} we have written the overall normalization 
as a real-space product for simplicity. It must also be intended as filtered: we will investigate this point 
in more detail in Section~\ref{subsec:covariance}. The higher-derivative stochasticities, 
i.e.~the higher orders in an expansion of the noise power spectrum in $\smash{R_\ast^2k^2}$ are under 
perturbative control in real space as long as $\Lambda < 1/R_\ast$ (for more details we refer to Section~4.3 of \cite{Cabass:2020nwf}). 

In the next section we will start from \eq{like_review-8} to derive an expression for the conditional likelihood in 
redshift space. It is important to emphasize that we will work in the infinite-$\Lambda$ limit. This is necessary if 
we want to achieve a nonperturbative expression for the redshift-space likelihood: 
indeed, all the manipulations we will make rely strongly on the transformation from the rest frame 
to redshift space being a local functional of the galaxy density in real space, cf.~\eq{notation-10}. We discuss how 
to connect to the results above (and by extension to \cite{Schmidt:2018bkr,Elsner:2019rql,2020arXiv200406707S}, 
where $\Lambda$ is kept finite), in Section~\ref{subsec:cutoffs}. 

Let us elaborate more on the infinite-$\Lambda$ limit. This must be intended only as a ``trick'' 
that allows us to compute exactly the contribution to the likelihood of the coordinate change to 
redshift space,~\eq{RS_review-1}. In this limit, there are many different contributions to the 
likelihood that also become important (the non-Gaussianity of the noise, the modulation of the 
noise by long-wavelength modes, etc.). By focusing only on the coordinate change in \eq{RS_review-1}, 
however, we are isolating at all orders in perturbations the terms that will not be affected by 
renormalization once we reintroduce a finite $\Lambda$ in Section~\ref{subsec:cutoffs}: even after 
short-scale modes are integrated out, the form of the ``displacement'' of the galaxy field to 
redshift space is unaffected (this is similar to what happens to the displacement from Lagrangian 
to Eulerian coordinates for the rest-frame likelihood, as discussed e.g.~in \cite{Schmidt:2018bkr,Elsner:2019rql,Cabass:2019lqx}). 

Before proceeding, we notice that in \cite{Cabass:2020nwf} we derived the impact of the stochasticities of bias coefficients 
on the conditional likelihood in the assumption that they follow a Gaussian probability distribution (on large scales this is 
the more relevant correction to a Gaussian likelihood with constant covariance). This effectively results in 
the replacement $\smash{P^{\{0\}}_{\eps_g}\to P_\eps[\delta]}$, where $P_\eps[\delta]$ is a positive-definite field-dependent 
covariance. We investigate how to include this in Section~\ref{subsec:zero_noise_normalization_etc}.

%********************************************************
\section{Main result}
\label{sec:main_result}
%********************************************************

\noindent First, let us define what we are after. For the purposes of Bayesian forward modeling, 
we still need the conditional likelihood for the data given the matter density and velocity in 
the rest frame. Indeed, this is what gravity-only N-body simulations most naturally output. 
The only difference with the conditional likelihood of Section~\ref{subsec:eft_likelihood_review} 
is that we now want to have the data in redshift space. Hence we need the likelihood 
\begin{equation}
\label{eq:what_we_need}
{\cal P}[\tilde{\delta}_g|\delta,\!\vec{v}]\,\,, 
\end{equation} 
which in terms of the joint likelihood $\smash{{\cal P}[\tilde{\delta}_g,\delta,\!\vec{v}]}$ and the likelihood 
for the matter density and velocity fields is given by 
\begin{equation}
\label{eq:what_we_need-joint_likelihood}
{\cal P}[\tilde{\delta}_g|\delta,\!\vec{v}] = \frac{{\cal P}[\tilde{\delta}_g,\delta,\!\vec{v}]} 
{{\cal P}[\delta,\!\vec{v}]}\,\,. 
\end{equation} 

Given that the relation between $\vec{v}_g$ and $\delta,\!\vec{v}$ is fully deterministic 
(its noise is discussed in Section~\ref{sec:velocity_noise}), we can compute the conditional likelihood of \eq{what_we_need} 
via the functional coordinate change summarized by \eqsIV{notation-3-B}{notation-9-A}{notation-9-B}{notation-9-C}. 
We do this in Section~\ref{subsec:coordinate_change}. 
Another way to arrive at $\smash{{\cal P}[\tilde{\delta}_g|\delta,\!\vec{v}]}$ 
is by integrating out the noise $\eps_g$, similarly to what we do 
with the rest-frame likelihood. This provides an additional check of our end 
result. We do this in Section~\ref{subsec:integrating_out_noise}. 

Finally, in Section~\ref{subsec:zero_noise_normalization_etc} we perform additional checks of the soundness 
of our result, and discuss how to include the modulation of the noise power spectrum by the matter field.

%********************************************************
\subsection{Calculation via functional coordinate change}
\label{subsec:coordinate_change}
%********************************************************

\noindent From \eqsIV{notation-3-B}{notation-9-A}{notation-9-B}{notation-9-C}, we see that the coordinate change we need to do is 
\begin{equation} 
\label{eq:coordinate_change_recall} 
\tilde{\delta}_g(\tilde{\vec{x}}) = \frac{1+\delta_g\big(\mathscr{R}(\tilde{\vec{x}})\big)} 
{\tilde{J}[\delta,\!\vec{v}](\tilde{\vec{x}})} - 1\,\,, 
\end{equation} 
whose inverse is 
\begin{equation} 
\label{eq:coordinate_change-1}
\delta_g(\vec{x}) = J[\delta,\!\vec{v}](\vec{x})\,\Big(1+\tilde{\delta}_g\big(\mathscr{R}^{-1}(\vec{x})\big)\Big) - 1 \,\,. 
\end{equation} 
Importantly, $\smash{\delta,\!\vec{v}}$ do not change. 

The fact that $\smash{\delta,\!\vec{v}}$ are not touched by the coordinate change has two consequences. First, 
thanks to \eq{what_we_need-joint_likelihood}, it tells us that the transformation of the conditional 
likelihood is the same as that of the joint likelihood. That is, $\smash{{\cal P}[\tilde{\delta}_g|\delta,\!\vec{v}]}$ is given by 
\begin{equation}
\label{eq:coordinate_change-2} 
\begin{split}
&\frac{1}{\abs[\bigg]{\dfrac 
{\partial(\tilde{\delta}_g,\delta,\!\vec{v})}
{\partial(\delta_g,\delta,\!\vec{v})}}}\, 
\Bigg(\prod_{\vec{x}}\sqrt{\frac{1}{2\pi P_{\eps_g}^{\{0\}}}}\,\Bigg)\exp\Bigg(
{-\frac{1}{2}}\int\dif^3x\,\frac{\big(\delta_g(\vec{x}) 
- \delta_{g,{\rm det}}[\delta](\vec{x})\big)^2}{P_{\eps_g}^{\{0\}}}\Bigg) \\
&\text{with $\delta_g(\vec{x}) = J[\delta,\!\vec{v}](\vec{x})\,\Big(1+\tilde{\delta}_g\big(\mathscr{R}^{-1}(\vec{x})\big)\Big) 
- 1$ in the exponent\,\,.} 
\end{split} 
\end{equation} 
Here the expression for the rest-frame likelihood is the one of \eq{like_review-8}, with 
$\delta_g$ in the exponent replaced with its expression in terms of $\smash{\tilde{\delta}_g}$ 
as per \eq{coordinate_change-1}, and the overall factor is the functional Jacobian of the coordinate change 
(for which we have used a notation that emphasizes how $\smash{\delta,\!\vec{v}}$ are not changed). 
Second, the continuum generalization of Leibniz's rule for determinants\footnote{I.e.~ 
\begin{equation}
\label{eq:leibniz_rule}
\det
\begingroup 
\begin{pmatrix} 
A & B \\ 
0 & D
\end{pmatrix}
\endgroup 
= 
\det A\,\det D 
= 
\det 
\begingroup 
\begin{pmatrix} 
A & 0 \\ 
C & D
\end{pmatrix}
\endgroup
\end{equation}
for any $n\times n$ matrix $A$ and $m\times m$ matrix $D$.} 
ensures that 
\begin{equation}
\label{eq:coordinate_change-3}
\frac{1}{\abs[\bigg]{\dfrac 
{\partial(\tilde{\delta}_g,\delta,\!\vec{v})}
{\partial(\delta_g,\delta,\!\vec{v})}}} = 
\abs[\bigg]{\frac{\partial(\delta_g,\delta,\!\vec{v})}{\partial(\tilde{\delta}_g,{\delta},\!{\vec{v}})}} 
= \abs[\bigg]{\frac{\partial\delta_g}{\partial\tilde{\delta}_g}} 
\abs[\bigg]{\frac{\partial(\delta,\!\vec{v})}{\partial({\delta},\!{\vec{v}})}}
= \abs[\bigg]{\frac{\partial\delta_g}{\partial\tilde{\delta}_g}}\,\,. 
\end{equation} 
This is because the Jacobian matrix is triangular in field space, thanks to the 
fact that $\smash{\delta,\!\vec{v}}$ do not depend on the galaxy field. 
For such matrices the determinant is the product of the determinants on the diagonal. 

Let us then compute the functional Jacobian of \eq{coordinate_change-3}. 
We only need the functional derivative 
\begin{equation}
\label{eq:coordinate_change-4}
\frac{\partial\delta_g(\vec{x})}{\partial\tilde{\delta}_g(\tilde{\vec{x}}')}\,\,. 
\end{equation} 
From \eq{coordinate_change-1}, together with \eq{notation-11}, we see that it is equal to 
\begin{equation}
\label{eq:coordinate_change-5}
J[\delta,\!\vec{v}](\vec{x})\,\delta^{(3)}_{\rm D}\big(\mathscr{R}^{-1}(\vec{x}) - \tilde{\vec{x}}'\big)\,\,. 
\end{equation} 
We can then use the property of the Dirac delta function (valid for any invertible function $f$) 
\begin{equation}
\label{eq:coordinate_change-6}
\delta^{(3)}_{\rm D}\big(f^{-1}(\vec{x}) - \tilde{\vec{x}}'\big) 
= \frac{\delta^{(3)}_{\rm D}(\vec{x}-\vec{x}')}{\abs[\bigg]{\dfrac{\partial f^{-1}(\vec{x})}{\partial\vec{x}}}} 
\end{equation} 
together with \eq{notation-2}, i.e.~ 
\begin{equation}
\label{eq:coordinate_change-7}
\abs[\bigg]{\frac{\partial \mathscr{R}^{-1}(\vec{x})}{\partial\vec{x}}} = J[\delta,\!\vec{v}](\vec{x})\,\,,
\end{equation}
to find 
\begin{equation}
\label{eq:coordinate_change-8}
\frac{\partial\delta_g(\vec{x})}{\partial\tilde{\delta}_g(\tilde{\vec{x}}')} 
= \delta^{(3)}_{\rm D}(\vec{x} - \vec{x}')\,\,, 
\end{equation} 
which implies 
\begin{equation}
\label{eq:coordinate_change-9}
\abs[\bigg]{\frac{\partial\delta_g}{\partial\tilde{\delta}_g}} = 1\,\,. 
\end{equation}

Switching to $\tilde{\vec{x}}$ in the exponent of \eq{coordinate_change-2}, we then find 
\begin{equation}
\label{eq:coordinate_change-10}
{\cal P}[\tilde{\delta}_g|\delta,\!\vec{v}] = 
\Bigg(\prod_{\vec{x}}\sqrt{\frac{1}{2\pi P_{\eps_g}^{\{0\}}}}\,\Bigg)\exp\Bigg(
{-\frac{1}{2}}\int\dif^3\tilde{x}\,\frac{\big( 
\tilde{\delta}_g(\tilde{\vec{x}}) - \tilde{\delta}_{g,{\rm det}}[\delta,\!\vec{v}](\tilde{\vec{x}})\big)^2} 
{P_{\eps_g}^{\{0\}}/\tilde{J}[\delta,\!\vec{v}](\tilde{\vec{x}})}\Bigg)\,\,, 
\end{equation} 
where the deterministic galaxy overdensity in redshift space is given by \eq{notation-10}, and 
the determinant $\smash{\tilde{J}[\delta,\!\vec{v}](\tilde{\vec{x}})}$ is the one of \eq{notation-9-C}.

%********************************************************
\subsection{Integrating out the noise}
\label{subsec:integrating_out_noise}
%********************************************************

\noindent In this section we arrive at the same result of \eq{coordinate_change-10} via another route, i.e.~by 
integrating out the noise $\eps_g(\vec{x})$ similarly to what we do to arrive at the rest-frame likelihood. 

Once we account for the noise $\eps_g(\vec{x})$, \eqsII{notation-3-B}{notation-10} tell us that the relation 
between $\smash{\tilde{\delta}_g(\tilde{\vec{x}})}$ and $\smash{\delta_{g,{\rm det}}[\delta](\vec{x})}$ is 
\begin{equation}
\label{eq:integrating_out_noise-1}
\tilde{\delta}_g(\tilde{\vec{x}}) = \tilde{\delta}_{g,{\rm det}}[\delta,\!\vec{v}](\tilde{\vec{x}}) + 
\frac{\eps_g\big(\mathscr{R}(\tilde{\vec{x}})\big)}{\tilde{J}[\delta,\!\vec{v}](\tilde{\vec{x}})}\,\,. 
\end{equation} 
Let us then multiply the Gaussian likelihood for $\eps_g(\vec{x})$, 
that for a constant noise power spectrum can be written as 
\begin{equation}
\label{eq:integrating_out_noise-2}
{\cal P}[\eps_g] = \Bigg(\prod_{\vec{x}}\sqrt{\frac{1}{2\pi P_{\eps_g}^{\{0\}}}}\,\Bigg) 
\exp\Bigg({-\frac{1}{2}}\int\dif^3x\,\frac{\eps^2_g(\vec{x})}{P^{\{0\}}_{\eps_g}}\Bigg)\,\,,
\end{equation} 
by a Dirac delta functional 
\begin{equation}
\label{eq:integrating_out_noise-3}
\delta^{(\infty)}_{\rm D}\Bigg(\tilde{\delta}_g(\tilde{\vec{x}}) 
- \tilde{\delta}_{g,{\rm det}}[\delta,\!\vec{v}](\tilde{\vec{x}}) - 
\frac{\eps_g\big(\mathscr{R}(\tilde{\vec{x}})\big)}{\tilde{J}[\delta,\!\vec{v}](\tilde{\vec{x}})}\Bigg) 
\end{equation} 
and functionally integrate over $\smash{{\cal D}\eps_g = \prod_{\vec{x}}\dif\eps_g(\vec{x})}$. To do this, let us define 
\begin{equation}
\label{eq:integrating_out_noise-4}
\tilde{\eps}_g(\tilde{\vec{x}}) = \frac{\eps_g\big(\mathscr{R}(\tilde{\vec{x}})\big)} 
{\tilde{J}[\delta,\!\vec{v}](\tilde{\vec{x}})}\,\,. 
\end{equation} 
The functional generalization of the change-of-coordinates rule for the Dirac 
delta functional makes \eq{integrating_out_noise-3} equal to 
\begin{equation}
\label{eq:integrating_out_noise-5}
\frac{1}{\abs[\bigg]{\dfrac{\partial\tilde{\eps}_g(\tilde{\vec{x}})}{\partial\eps_g(\vec{x}')}}}\, 
\delta^{(\infty)}_{\rm D}\Big(\eps_g(\vec{x}) + \big\{1 + \delta_{g,{\rm det}}[\delta](\vec{x})\big\} 
- J[\delta,\!\vec{v}](\vec{x})\big\{1+\tilde{\delta}_g\big(\mathscr{R}^{-1}(\vec{x})\big)\big\}\Big)\,\,. 
\end{equation} 
The overall Jacobian is straightforward to evaluate. From \eq{integrating_out_noise-4} 
we need to evaluate the functional determinant of 
\begin{equation}
\label{eq:integrating_out_noise-6}
\frac{\delta^{(3)}_{\rm D}\big(\mathscr{R}(\tilde{\vec{x}}) - \vec{x}'\big)} 
{\tilde{J}[\delta,\!\vec{v}](\tilde{\vec{x}})}\,\,, 
\end{equation} 
which is equal to $\delta^{(3)}_{\rm D}(\tilde{\vec{x}}-\tilde{\vec{x}}')$ thanks to \eq{notation-3-A} 
and the properties of the three-dimensional Dirac delta function. Hence we have 
\begin{equation}
\label{eq:integrating_out_noise-7}
\abs[\bigg]{\dfrac{\partial\tilde{\eps}_g(\tilde{\vec{x}})}{\partial\eps_g(\vec{x}')}} = 1\,\,. 
\end{equation}
Integrating $\eps_g(\vec{x})$ out via \eqsII{integrating_out_noise-5}{integrating_out_noise-7} then gives 
\begin{equation}
\label{eq:integrating_out_noise-8} 
\begin{split}
{\cal P}[\tilde{\delta}_g|\delta,\!\vec{v}] &= 
\Bigg(\prod_{\vec{x}}\sqrt{\frac{1}{2\pi P_{\eps_g}^{\{0\}}}}\,\Bigg) \\
&\;\;\;\;\times\exp\Bigg({-\frac{1}{2}}\int\dif^3x\,\frac{\big\{1 + 
\tilde{\delta}_g\big(\mathscr{R}^{-1}(\vec{x})\big) 
- \big(1+\delta_{g,{\rm det}}[\delta](\vec{x})\big)/J[\delta,\!\vec{v}](\vec{x})\big\}^2} 
{P_{\eps_g}^{\{0\}}/J^2[\delta,\!\vec{v}](\vec{x})}\Bigg)\,\,. 
\end{split} 
\end{equation} 
Finally, switching to redshift space in the integral in the exponent we obtain the same result as in \eq{coordinate_change-10}. 
In Appendix~\ref{app:integrating_out_noise_appendix} we confirm this result a final time via manipulations very similar to the above.

%********************************************************
\subsection{Limit of zero noise, normalization and stochasticity of bias coefficients}
\label{subsec:zero_noise_normalization_etc}
%********************************************************

\noindent In this section we look in more detail at the result of \eqsII{intro}{coordinate_change-10}. 
\begin{itemize}[leftmargin=*]
\item First, we confirm that in the limit $\smash{P^{\{0\}}_{\eps_g}\to 0}$ we obtain a Dirac delta functional that sets 
$\smash{\tilde{\delta}_g(\tilde{\vec{x}})}$ equal to its deterministic expression of \eq{notation-10}. 
\item Related to this, we check that our likelihood has the correct normalization with respect to 
$\smash{\tilde{\delta}_g(\tilde{\vec{x}})}$. That is, we check that integrating it over 
$\smash{\tilde{\delta}_g(\tilde{\vec{x}})}$ gives $1$. As a consequence, 
we prove that indeed \eq{coordinate_change-10} is equal to \eq{intro}. 
There is also an important advantage in moving from \eq{coordinate_change-10} to \eq{intro}, 
since in the latter the likelihood is written exclusively in redshift-space coordinates 
$\smash{\tilde{\vec{x}}}$, that are the coordinates used in observations. 
\item Finally, we discuss how to include the impact of the stochasticity of bias coefficients following \cite{Cabass:2020nwf}, 
that gives the leading correction to the rest-frame likelihood with constant noise power spectrum. 
\end{itemize}

%********************************************************
\subsubsection*{Limit of zero noise}
\label{subsubsec:zero_noise_limit}
%********************************************************

\noindent Consider the probability distribution of $\eps_g(\vec{x})$, i.e.~\eq{integrating_out_noise-2}. 
In the limit $\smash{P^{\{0\}}_{\eps_g}\to 0}$ it is equal to a Dirac delta functional that sets $\eps_g(\vec{x})$ 
identically equal to zero. Hence, if we multiply it by \eq{integrating_out_noise-3} and integrate over the noise, we must obtain 
\begin{equation}
\label{eq:zero_noise_limit-1}
\delta^{(\infty)}_{\rm D}\big(\tilde{\delta}_g(\tilde{\vec{x}}) 
- \tilde{\delta}_{g,{\rm det}}[\delta,\!\vec{v}](\tilde{\vec{x}})\big)\,\,, 
\end{equation} 
where the deterministic galaxy overdensity in redshift space is defined by \eq{notation-10}. 

To see how this works out we take the limit $\smash{P^{\{0\}}_{\eps_g}\to 0}$ in \eq{integrating_out_noise-8}. We obtain 
\begin{equation}
\label{eq:zero_noise_limit-2}
\Bigg(\prod_{\vec{x}}\frac{1}{J[\delta,\!\vec{v}](\vec{x})}\Bigg)\, 
\delta^{(\infty)}_{\rm D}\Big(\tilde{\delta}_g\big(\mathscr{R}^{-1}({\vec{x}})\big) - 
\tilde{\delta}_{g,{\rm det}}[\delta,\!\vec{v}]\big(\mathscr{R}^{-1}({\vec{x}})\big)\Big)\,\,. 
\end{equation} 
We can then use the properties of the Dirac delta functional to change the argument of the two fields. 
The overall factor we get from the change of variables is 
\begin{equation}
\label{eq:zero_noise_limit-3} 
\dfrac{1}{\abs[\big]{\delta^{(3)}_{\rm D}\big(\mathscr{R}^{-1}({\vec{x}}) - \tilde{\vec{x}}'\big)}} 
\end{equation} 
where, as in all the equations above, we denote the determinant by $\smash{\abs{\cdot}}$ 
(here the determinant of a matrix in $\vec{x},\vec{x}'$). \eq{zero_noise_limit-3} is equal to 
\begin{equation} 
\label{eq:zero_noise_limit-4}
\Bigg(\prod_{\vec{x}}\frac{1}{J[\delta,\!\vec{v}](\vec{x})}\Bigg)^{-1} 
\end{equation} 
thanks to \eq{notation-2} and the properties of the three-dimensional Dirac delta function, hence confirming \eq{zero_noise_limit-1}.

%********************************************************
\subsubsection*{Normalization of the likelihood}
\label{subsubsec:normalization}
%********************************************************

\noindent With similar manipulations we can show that the redshift-space conditional likelihood is normalized if we integrate 
over the data, i.e.~in $\smash{{\cal D}\tilde{\delta}_g = \prod_{\tilde{\vec{x}}}\dif\tilde{\delta}_g(\tilde{\vec{x}})}$. 
In order to do this we switch the argument of the fields in the likelihood back to $\vec{x}$, that is we take 
\eq{integrating_out_noise-8}, and perform the corresponding change of variables in the functional measure. We define 
\begin{equation}
\label{eq:normalization-1}
\tilde{\delta}_g\big(\mathscr{R}^{-1}(\vec{x})\big) = \Delta(\vec{x})\,\,. 
\end{equation} 
Thanks to the Jacobian of the inverse being the inverse of the Jacobian, the functional measure changes into 
\begin{equation}
\label{eq:normalization-2}
{\cal D}\tilde{\delta}_g = {\cal D}{\Delta}\,\dfrac{1}{\abs[\bigg]{\dfrac{\partial\Delta(\vec{x})} 
{\partial\tilde{\delta}_g(\tilde{\vec{x}}')}}}\,\,,
\end{equation} 
where $\smash{{\cal D}{\Delta} = \prod_{\vec{x}}\dif\Delta(\vec{x})}$ and 
\begin{equation}
\label{eq:normalization-3}
\abs[\bigg]{\dfrac{\partial \Delta(\vec{x})}{\partial\tilde{\delta}_g(\tilde{\vec{x}}')}} = 
\abs[\big]{\delta^{(3)}_{\rm D}(\mathscr{R}^{-1}(\vec{x}) - \tilde{\vec{x}}')} = 
\prod_{\vec{x}}\frac{1}{J[\delta,\!\vec{v}](\vec{x})}\,\,, 
\end{equation} 
as in \eqsII{zero_noise_limit-3}{zero_noise_limit-4}. 

Hence, combining \eqsII{normalization-2}{normalization-3} with \eq{integrating_out_noise-8}, 
more precisely the fact that the overall factor multiplying the exponential is now 
\begin{equation}
\label{eq:normalization-4}
\prod_{\vec{x}}\sqrt{\frac{J^2[\delta,\!\vec{v}](\vec{x})}{2\pi P_{\eps_g}^{\{0\}}}}\,\,,
\end{equation}
after a simple shift of integration variables we see that the integral in $\smash{{\cal D}\Delta}$ is equal to $1$. 

It is then useful to recast our result in a form that makes more clear what is the 
limit of zero noise and the normalization of the likelihood. This just amounts to writing 
the overall normalization as a product over redshift-space coordinates as in \eq{intro}, i.e.~ 
\begin{equation}
\label{eq:normalization-5}
{\cal P}[\tilde{\delta}_g|\delta,\!\vec{v}] = \Bigg(\prod_{\tilde{\vec{x}}}\sqrt{ 
\frac{\tilde{J}[\delta,\!\vec{v}](\tilde{\vec{x}})}{2\pi P_{\eps_g}^{\{0\}}}}\,\Bigg)\exp\Bigg( 
{-\frac{1}{2}}\int\dif^3\tilde{x}\,\frac{\big( 
\tilde{\delta}_g(\tilde{\vec{x}}) - \tilde{\delta}_{g,{\rm det}}[\delta,\!\vec{v}](\tilde{\vec{x}})\big)^2} 
{P_{\eps_g}^{\{0\}}/\tilde{J}[\delta,\!\vec{v}](\tilde{\vec{x}})}\Bigg)\,\,. 
\end{equation}

%********************************************************
\subsubsection*{Stochasticity of bias coefficients}
\label{subsubsec:stochasticity_bias_coefficients}
%********************************************************

\noindent All of the manipulations of this section go through in the same way if 
the constant noise power spectrum is replaced by a field-dependent one, i.e.~ 
\begin{equation} 
\label{eq:stochasticity_bias_coefficients-1} 
P^{\{0\}}_{\eps_g}\to P_\eps[\delta]\,\,. 
\end{equation} 
This is (at leading order in derivatives) the effect that the stochasticity of bias coefficients, which 
describes the fact that the noise generated by integrating short-scale modes can be modulated by 
long-wavelength perturbations, has on the rest-frame likelihood \cite{Cabass:2020nwf}. More precisely we have 
\begin{equation}
\label{eq:stochasticity_bias_coefficients-2}
P_{\eps}[\delta](\vec{x}) = P^{\{0\}}_{\eps_g} 
+ 2\sum_{O}P_{\varepsilon_g\varepsilon_{g,{O}}}^{\{0\}} 
{O}[\delta](\vec{x}) 
+ \sum_{{O},{O}'}P_{\varepsilon_{g,{O}}\varepsilon_{g,{O}'}}^{\{0\}} 
{O}[\delta](\vec{x}){O}'[\delta](\vec{x})\,\,, 
\end{equation}
where the bias coefficients $b_O$ and the operators $O[\delta]$ are defined in \eq{notation-5}, 
and $\smash{P_\eps[\delta]\geq 0}$ as long as the power spectra of the noises are such that the matrix 
\begin{equation}
\label{eq:stochasticity_bias_coefficients-3}
\begingroup 
\setlength\arraycolsep{3pt}
\begin{pmatrix} 
P^{\{0\}}_{\eps_g} & P_{\varepsilon_g\varepsilon_{g,{\delta}}}^{\{0\}} & 
\cdots & P_{\varepsilon_g\varepsilon_{g,{O}}}^{\{0\}} & \cdots \\
P_{\varepsilon_g\varepsilon_{g,{\delta}}}^{\{0\}} & P^{\{0\}}_{\eps_{g,\delta}} 
& \cdots & P_{\varepsilon_{g,{\delta}}\varepsilon_{g,O}}^{\{0\}} & \cdots \\
\vdots & \vdots & \ddots & \cdots & \cdots \\
P_{\varepsilon_g\varepsilon_{g,{O}}}^{\{0\}} & P_{\varepsilon_{g,{\delta}}\varepsilon_{g,O}}^{\{0\}} 
& \vdots & P^{\{0\}}_{\eps_{g,O}} & \cdots \\ 
\vdots & \vdots & \vdots & \vdots & \ddots \\ 
\end{pmatrix}
\endgroup 
\end{equation} 
is positive-definite (the fields $\smash{\eps_{g,O}}$ describe the stochasticity of the bias 
coefficients via the shift $\smash{b_O\to b_O+\eps_{g,O}}$ in the deterministic bias expansion. 
For simplicity of notation, and following \cite{Desjacques:2016bnm}, we have used the shorthand 
$\smash{P_{\varepsilon_{g,{O}}\varepsilon_{g,O}}^{\{0\}}} = P_{\varepsilon_{g,O}}^{\{0\}}$). 

The replacement of \eq{stochasticity_bias_coefficients-1} leads, then, to the 
final expression for the conditional likelihood in redshift space, i.e.~ 
\begin{equation}
\label{eq:stochasticity_bias_coefficients-4}
{\cal P}[\tilde{\delta}_g|\delta,\!\vec{v}] = \Bigg(\prod_{\tilde{\vec{x}}}\sqrt{ 
\frac{\tilde{J}[\delta,\!\vec{v}](\tilde{\vec{x}})}{2\pi \tilde{P}_\eps[\delta,\!\vec{v}](\tilde{\vec{x}})}}\,\Bigg) 
\exp\Bigg({-\frac{1}{2}}\int\dif^3\tilde{x}\,\frac{\big( 
\tilde{\delta}_g(\tilde{\vec{x}}) - \tilde{\delta}_{g,{\rm det}}[\delta,\!\vec{v}](\tilde{\vec{x}})\big)^2} 
{\tilde{P}_\eps[\delta,\!\vec{v}](\tilde{\vec{x}})/\tilde{J}[\delta,\!\vec{v}](\tilde{\vec{x}})}\Bigg)\,\,, 
\end{equation} 
where we have defined 
\begin{equation}
\label{eq:stochasticity_bias_coefficients-5}
\tilde{P}_\eps[\delta,\!\vec{v}](\tilde{\vec{x}}) = P_\eps[\delta]\big(\mathscr{R}(\tilde{\vec{x}})\big)\,\,, 
\end{equation} 
with the additional dependence on $\vec{v}$ coming from the change of the argument, cf.~\eq{notation-9-A}.

%********************************************************
\section{Cutoffs and perturbativity}
\label{sec:cutoffs_and_perturbativity}
%********************************************************

\noindent So far we have worked in the infinite-$\Lambda$ limit. As discussed at the end of Section~\ref{subsec:eft_likelihood_review}, 
this was necessary to arrive at an expression for the likelihood valid at all orders in perturbation theory. In turn, this is 
necessary if one wants to sample the likelihood numerically: perturbative (Edgeworth-like) expansions of the likelihood are 
not suitable to numerical manipulations. Let us now discuss how a finite $\Lambda$ can be reinstated and, most importantly, 
what are the consequences.

%********************************************************
\subsection{Reintroducing the cutoff}
\label{subsec:cutoffs}
%********************************************************

\noindent In Section~\ref{subsec:eft_likelihood_review} we have seen that, for the Gaussian likelihood 
with constant noise power spectrum of \cite{Schmidt:2018bkr,Elsner:2019rql,Cabass:2019lqx}, the cutoff 
was implemented in the following way: 
\begin{itemize}[leftmargin=*]
\item we construct the deterministic bias expansion $\delta_{g,{\rm det}}$ from a matter field $\delta$ 
cut at a scale $\Lambda$. After getting $\delta$ from, e.g., an N-body simulation, one can transform it 
to Fourier space, cut it at $\Lambda$, transform it back to real space and construct the (local in real space) 
deterministic bias expansion out of it (Refs.~\cite{Schmidt:2018bkr,Elsner:2019rql,2020arXiv200406707S} follow this procedure); 
\item once we have the real-space $\delta_{g,{\rm det}}[\delta_\Lambda]$, we can transform it to Fourier space, cut it, 
and then transform it back to real space; 
\item the data $\delta_g$ are also cut at $\Lambda$ following the same procedure. 
\end{itemize} 
The exponent in the likelihood, then, is constructed by taking the difference between the real-space 
$\smash{\delta_{g,\Lambda}}$ and $\smash{\delta_{g,{\rm det},\Lambda}[\delta_\Lambda]}$ constructed as above, squaring it, 
dividing it by $\smash{{-2}P^{\{0\}}_{\eps_g}}$ and integrating in $\smash{\dif^3x}$. 

To understand how this generalizes to our redshift-space likelihood we can first recall 
how the cutoff $\Lambda$ arises when we connect the likelihood to correlation functions. 
Roughly speaking, when we want to describe correlation functions via effective field theory techniques we 
deal with two types of momenta: loop momenta and external momenta. Loop momenta are integrated 
over. They run to infinity and the resulting UV-sensitivity of diagrams is absorbed by counterterms. 
Even after we have renormalized the theory, however, we still do not want to take external momenta 
to be arbitrarily hard. We know that our effective field theory fails around some physical scale (the 
nonlinear scale $k_{\rm NL}$, or the galaxy nonlocality scale $1/R_\ast$), so we cannot trust our predictions 
on scales shorter than that. Hence, we take external momenta to be smaller than some $\Lambda$: the larger 
the $\Lambda$, the more important the contribution of higher-order operators will be. 

The presence of $\Lambda$ can be accounted for straightforwardly once we collect all correlation functions in 
the so-called generating functional $Z$. This is a functional of an ``external current'' $J$ (not to be confused 
with \eq{notation-2}, i.e.~the Jacobian of the transformation from redshift space to the rest-frame coordinates), 
such that functional derivatives of $\ln Z$ evaluated at $J=0$ yield the connected correlation functions of the theory. 
If we take $J$ such that its Fourier modes are zero above $\Lambda$, we are then sure that we only 
compute correlation functions with external momenta softer than $\Lambda$. 

This has a direct consequence on the likelihood. Indeed, the likelihood is the Fourier transform of 
the generating functional, and the fields $\delta_g$ and $\delta$ are the ``duals'' to the currents 
$J_g$ and $J$ for the correlation functions of galaxies and matter, respectively. Since these currents 
are cut at $\Lambda$, $\delta_g$ and $\delta$ are cut as well. Finally, the deterministic bias expansion 
$\smash{\delta_{g,{\rm det}}}$ is also cut because it is linearly coupled to the current $J_g$ in 
the functional integral that defines the generating functional 
\cite{Carroll:2013oxa,Cabass:2019lqx,Cabass:2020nwf}.\footnote{One can schematically picture this in the following way. 
The generating functional for galaxies can be defined as a functional integral over the matter field $\delta$, i.e.~ 
\begin{equation}
\label{eq:linear_mixing}
Z[J_g] = \int{\cal D}\delta\,{\cal P}[\delta] \exp\bigg(\int\dif^3x\,
J_g(\vec{x})\delta_{g,{\rm det}}[\delta](\vec{x}) + \cdots\bigg)\,\,, 
\end{equation} 
where ${\cal P}[\delta]$ is the matter probability distribution and $\cdots$ denotes terms of higher order in $J_g$ 
that describe the noise (its power spectrum, its higher-order correlation functions, and the stochasticity of bias coefficients), 
entering as counterterms to absorb the UV dependencies that arise once we integrate out the short modes of $\delta$. 
Importantly, the deterministic bias expansion is coupled linearly to the current. Hence, if the current is cut at $\Lambda$, 
modes of the field $\smash{\delta_{g,{\rm det}}[\delta](\vec{x})}$ above $\Lambda$ do not appear in the path integral above. 
For more details we refer to \cite{Cabass:2019lqx,Cabass:2020nwf}. See also \cite{Carroll:2013oxa} for a discussion in 
the context of the EFTofLSS for matter.} 

The exact same reasoning can then be applied to our redshift-space conditional likelihood. 
The Fourier modes, which are defined by the Fourier transform over the cartesian coordinates $\smash{\tilde{\vec{x}}}$, 
of the data $\smash{\tilde{\delta}_g}$ and of the deterministic bias expansion in redshift space 
$\smash{\tilde{\delta}_{g,{\rm det}}}$ are set to zero above a cutoff $\smash{\Lambda}$. The same 
happens for the Fourier modes (defined via Fourier transform over $\smash{\vec{x}}$) of the matter field. 
Importantly, the covariance $\smash{P_{\eps_g}^{\{0\}}/\tilde{J}[\delta,\!\vec{v}](\tilde{\vec{x}})}$ 
in the exponent of \eqsII{intro}{normalization-5} (or their analog in \eq{stochasticity_bias_coefficients-4} 
when we want to include the stochasticity of bias coefficients), is constructed from the filtered matter field. 
The integral is then carried out over $\smash{\dif^3\tilde{x}}$. 
\begin{itemize}[leftmargin=*] 
\item The fact that the integral is constructed from filtered fields is what makes the connection with 
a perturbative treatment possible. This is discussed in Section~\ref{subsec:perturbativity} below. 
\item The numerator in \eqsII{intro}{normalization-5} or \eq{stochasticity_bias_coefficients-4} 
is the square of a difference, hence it is manifestly positive. As long as the denominator is positive, 
then, the whole exponent is positive. Since we are dealing with multi-dimensional integrals this does not 
guarantee, however, that the likelihood is well-behaved. We come back to this in Section~\ref{subsec:covariance}. 
\item As a ``corollary'' of this discussion, we see how the possibility of working with filters different 
from an isotropic hard cut in Fourier space arises. Let us go back to correlation functions, and work in 
the distant-observer/flat-sky approximation (as it is usually done when discussing the EFTofLSS in redshift space, 
see e.g.~\cite{Perko:2016puo,Desjacques:2018pfv}). For the sake of this discussion 
let us also focus on the power spectrum. 
We then have at least two possibilities. Either we work by taking multipoles of the power spectrum 
(in some orthonormal basis), or we bin it in both $k = \abs{\vec{k}}$ and $\mu = \vers{k}\cdot\vers{n}$. 
The second one is the ``wedges'' approach proposed in \cite{Kazin:2011xt}. 
When translated in terms of cutoffs, it means that we can implement two different cuts: 
one on $k$ and one on $k_\parallel = \vec{k}\cdot\vers{n}$. The advantage of implementing a cut on $k_\parallel$ softer than 
the one on $k$ is that we can have more control on higher-order corrections from $u_\parallel$ 
(as we will see in a moment, any occurrence of $u_\parallel$ comes with a $\partial_\parallel$). 
Since the cutoffs that we implement in the analysis of correlation functions are exactly those 
on the fields appearing in the likelihood, we see that it is possible to cut 
$\smash{\tilde{\delta}_g}$ and $\smash{\tilde{\delta}_{g,{\rm det}}}$ via an anisotropic filter. 
We leave a more detailed investigation to future work. 
\end{itemize} 

So far we have not discussed how to deal with the overall factor in front of the exponential. 
In the real-space expression for the rest-frame likelihood with constant noise power spectrum, \eq{like_review-8}, 
it was intended as filtered at a scale $\Lambda$ as well. However, it was very simple to implement such filter there 
even if we worked in real space, thanks to the fact that $\smash{P^{\{0\}}_{\eps_g}}$ is constant. 
How do we proceed now that the covariance depends on $\vec{x}$ (or, better, the redshift-space position $\tilde{\vec{x}}$)? 
We come back to this important point in Section~\ref{subsec:covariance} below.

%********************************************************
\subsection{Connection with the perturbative treatment}
\label{subsec:perturbativity}
%********************************************************

\noindent We now have the tools to study the connection with the perturbative treatment of the EFTofLSS. 
The presence of the cutoffs is fundamental for this purpose. 
\begin{itemize}[leftmargin=*] 
\item The matter field is cut at $\Lambda$. This means that the real-space matter field is a collection of 
modes with wavenumbers up to $\abs{\vec{k}}=\Lambda$. Since we take $\Lambda$ below $k_{\rm NL}$, we know it is the 
linear power spectrum that controls the amplitude of these modes. For a power-law dimensionless linear power spectrum 
we have $\smash{\sqrt{\Delta^2(k)} = (k/k_{\rm NL})^{(3+n_\delta)/2}}$, where $\smash{\Delta^2(k) = 
k^3 P_{\rm L}(k)/2\pi^2}$. Then, the typical size of a filtered perturbation $\smash{\delta(\vec{x})}$ 
(in this section we drop the subscript ``$\Lambda$'' for simplicity of notation) scales as 
\begin{equation}
\label{eq:perturbativity-1}
\delta(\vec{x})\sim\bigg(\frac{\Lambda}{k_{\rm NL}}\bigg)^{\frac{3+n_\delta}{2}}\,\,. 
\end{equation} 
Taking $n$ derivatives of this field increases the scaling by $n$ powers of $\Lambda$. 
The index $n_\delta$ is between ${-2}$ and ${-1.5}$ on the scales where loop corrections in the EFTofLSS become important, 
see e.g.~\cite{Pajer:2013jj,Abolhasani:2015mra,Baldauf:2016sjb} and Section~4.1 of \cite{Desjacques:2016bnm}. 
Hence, as long as $\Lambda$ is sufficiently lower than $k_{\rm NL}$, \eq{perturbativity-1} ensures that 
$\smash{\abs{\delta(\vec{x})} < 1}$ and that $\smash{\abs{\vec{\nabla}^n\delta(\vec{x})/k_{\rm NL}^n} < 1}$. 
\item Consequently, in the deterministic bias expansion $\smash{\tilde{\delta}_{g,{\rm det}}}$ for the galaxy overdensity 
in redshift space the same perturbative expansion that we are used to when dealing with correlation functions applies. 
Let us see how this works. As shown e.g.~in \cite{Scoccimarro:2004tg} (see also Eq.~(9.44) of \cite{Desjacques:2016bnm}), 
after taking into account the change of argument from $\smash{\vec{x}}$ to $\smash{\tilde{\vec{x}}}$ we obtain 
\begin{equation}
\label{eq:perturbativity-2}
\tilde{\delta}_{g,{\rm det}} = \delta_{g,{\rm det}} + \sum_{n=1}^{+\infty}\frac{(-1)^n}{n!}\partial^n_\parallel 
\Big(u_\parallel^n(1+\delta_{g,{\rm det}})\Big)\,\,,
\end{equation} 
where all fields are evaluated at the redshift-space coordinate $\smash{\tilde{\vec{x}}}$ and 
$\smash{\delta_{g,{\rm det}}}$ is given by the bias expansion of \eq{notation-5}, constructed from a 
filtered matter field. The galaxy velocity $u_\parallel$ is given by its deterministic 
bias expansion, e.g.~\eqsII{notation-7}{notation-8}, constructed from a filtered $\delta$ and a filtered matter velocity field. 
\eq{perturbativity-1} ensures that a perturbative expansion of $\delta_{g,{\rm det}}[\delta_\Lambda]$ is under control. 
What about the new terms, i.e.~those involving $u_\parallel$? These are also under perturbative control. 
In fact, we see that $u_\parallel$ never appears by itself, but it always comes with a derivative $\partial_\parallel$ (be it 
acting on $u_\parallel$ itself or on other fields). Given that on large scales the matter velocity scales like 
$\smash{(\vec{\nabla}/\nabla^2)\delta}$, the additional derivative $\partial_\parallel$ ensures that any occurrence 
of $u_\parallel$ in \eq{perturbativity-2} scales with $\Lambda$ in the same way as in \eq{perturbativity-1}. Most importantly, 
recall that $\smash{\vec{v}}$ is actually proportional via a dimensionless coefficient to 
$\smash{({\cal H}\vec{\nabla}/\nabla^2)\delta}$. The factors of $\smash{{\cal H}}$ then simplify in \eq{RS_review-2}, and we get 
\begin{equation}
\label{eq:perturbativity_check_scale}
\partial_\parallel u_\parallel\sim u_\parallel\partial_\parallel\sim\bigg(\frac{\Lambda}{k_{\rm NL}}\bigg)^{\frac{3+n_\delta}{2}}\,\,. 
\end{equation} 
That is, the scaling with $\Lambda$ is still controlled by the nonlinear scale $k_{\rm NL}$ only.\footnote{Notice 
that here we are not considering the overall growth rate $f = \dif\ln D_1/\dif\ln a$ that $u_\parallel$ is proportional to 
at linear order in perturbations (equal to $1$ in an Einstein-de Sitter universe). Importantly, its presence does not 
affect the scaling of \eq{perturbativity_check_scale} with $\Lambda$.} 
\item The difference between the filtered $\smash{\tilde{\delta}_g}$ and $\smash{\tilde{\delta}_{g,{\rm det}}}$ is 
controlled by the rest-frame noise cut at $\Lambda$. We are justified in neglecting the transformation of the noise field 
to redshift space (given by \eq{integrating_out_noise-4}, effectively) because we have shown above that such transformation 
is under perturbative control. The same argument applies to the modulation of the noise power spectrum by the matter field, 
\eq{stochasticity_bias_coefficients-2}. Hence, the size of a noise fluctuation in real space scales as 
\begin{equation} 
\label{eq:perturbativity-3}
\sqrt{P^{\{0\}}_{\eps_g}\Lambda^3}\,\,. 
\end{equation} 
That is, it is controlled by the mean separation between tracers for $\smash{P^{\{0\}}_{\eps_g} = \alpha/\widebar{n}_g}$. 
This scaling is especially important because, at variance with the 
deterministic bias expansion, our likelihood does not include the non-Gaussianity of the noise at 
all orders. However, since the corrections from the noise non-Gaussianity come in the form of 
higher powers in $\smash{\tilde{\delta}_g - \tilde{\delta}_{g,{\rm det}}}$ \cite{Cabass:2019lqx}, 
we know that they are under control as long as $\Lambda$ is softer than $\smash{(1/\widebar{n}_g)^{1/3}}$. 
\item Finally, we can estimate the relative importance of the different ingredients in the likelihood of 
\eqsII{intro}{normalization-5}. More precisely, we can compare the relative importance of higher-order terms 
in the deterministic bias expansion with higher-order terms in the field-dependent covariance 
$\smash{P_{\eps_g}^{\{0\}}/\tilde{J}[\delta,\!\vec{v}](\tilde{\vec{x}})}$. 
Expanding the both the numerator and the denominator of the exponential around linear theory, we see 
that going up to $n$th order in perturbations in $\smash{\tilde{\delta}_{g,{\rm det}}}$ 
is always more important than going up to the same order in the covariance. 
The former is enhanced with respect to the latter by the ratio (recall that ${-2}\lesssim n_\delta\lesssim{-1.5}$) 
\begin{equation}
\label{eq:perturbativity-4}
\frac{\bigg(\dfrac{\Lambda}{k_{\rm NL}}\bigg)^{\frac{3+n_\delta}{2}}}{\sqrt{P^{\{0\}}_{\eps_g}\Lambda^3}} 
=\bigg(\frac{1}{k_{\rm NL}P^{\{0\}}_{\eps_g}}\bigg)^{\frac{3}{2}}\bigg(\frac{\Lambda}{k_{\rm NL}}\bigg)^{\frac{n_\delta}{2}}\,\,. 
\end{equation}
The same counting goes through when including the stochasticity of bias coefficients 
(\eqsII{stochasticity_bias_coefficients-4}{stochasticity_bias_coefficients-5} 
being the resulting likelihood), as discussed in \cite{Cabass:2020nwf}. 
\end{itemize} 

Before proceeding, it is important to emphasize that these conclusions rely on the 
assumption of negative $\smash{n_\delta}$. This assumption holds on mildly-nonlinear scales, where the 
EFTofLSS is usually employed to push the reach of perturbation theory. The linear power spectrum, 
however, turns around near $\smash{0.02\,h\,{\rm Mpc}^{-1}}$. On these large, linear scales, 
the relative importance of the various terms in the likelihood is changed: see e.g.~Fig.~\ref{fig:scalings} 
(this has been discussed also in \cite{Cabass:2019lqx,2020arXiv200406707S}).\footnote{Let us consider 
for example the relative importance of the noise non-Gaussianity and the contributions from the clustering bispectrum. 
In the deeply linear regime, it is not important which one of the two dominates: both are subleading with respect 
to the simple linear contribution $\smash{\delta_g - b_1\delta}$ in the conditional likelihood. 
The non-Gaussianity of the noise is still suppressed by \eq{noise_scaling}, while the effect of clustering is 
suppressed as $\smash{\Lambda^{3/2+n_\delta}}$, for positive $\smash{n_\delta}$.} 

The last of the bullet points above also allows us to address the positive-definiteness of the 
Jacobian $\smash{\tilde{J}}$ of the coordinate change to redshift space. As long as the cutoff 
$\Lambda$ is soft enough, \eq{notation-2} tells us that the Jacobian is equal to $1$ plus small corrections. 
It is important to emphasize, however, that there is no physical reason why this should hold 
as we take $\Lambda$ to be larger and larger. Indeed, it is possible that a simply connected 
volume in observed coordinates does not correspond to a simply connected region in rest-frame coordinates, i.e.~that 
the coordinate change $\mathscr{R}$ is multi-valued: this can happen for galaxies 
that have large peculiar velocities. Restricting $\Lambda$ to lie in the perturbative regime is what allows 
to get around this issue (see also Section~9.3.2 of \cite{Desjacques:2016bnm} for a discussion). 

We can compare and contrast this with the positive-definiteness of $\smash{P^{\{0\}}_{\eps_g}}$, 
or in general of $\smash{P_\eps[\delta]}$ in \eqsII{stochasticity_bias_coefficients-1}{stochasticity_bias_coefficients-2}. 
Independently of whether we construct $\smash{P_\eps[\delta]}$ from a filtered matter field or not, we know that it cannot 
be negative if the (Gaussian) noise likelihood has a positive-definite covariance. Consider the manifestly nonnegative combination 
\begin{equation}
\label{eq:perturbativity-5}
\Bigg(\eps_g(\vec{x})+\sum_O\eps_{g,O}(\vec{x})\,O[\delta](\vec{x})\Bigg)^2\,\,, 
\end{equation}
and functionally integrate it over the noise fields. As long as the noise covariance of \eq{stochasticity_bias_coefficients-3} 
is a positive-definite matrix, this integral is well defined and gives exactly \eq{stochasticity_bias_coefficients-2} 
times $\smash{\delta^{(3)}_{\rm D}(\vec{0})}$. Hence we know that also the combination in \eq{stochasticity_bias_coefficients-2} 
cannot be negative.\footnote{Notice that the proof goes through in the same way even if we cut the noise fields at some finite scale 
$\Lambda$. The only difference would be that $\smash{\delta^{(3)}_{\rm D}(\vec{0})}$ is replaced by its finite-$\Lambda$ counterpart 
evaluated at zero spatial separation, which is a positive number $\propto\Lambda^3$.} 
If we take a soft $\Lambda$, such that the higher-order terms in \eq{stochasticity_bias_coefficients-2} are 
small corrections to $\smash{P^{\{0\}}_{\eps_g}}$, we conclude that $\smash{P_\eps[\delta]}$ is manifestly positive.

%********************************************************
\subsection[Covariance and normalization]{Covariance and normalization\footnote{It is a pleasure to 
thank Fabian Schmidt for very useful discussions regarding the subject of this section.}}
\label{subsec:covariance}
%********************************************************

\noindent We have seen in the previous section that the presence of a cutoff 
allows to make the connection with the perturbative treatment manifest. 
In this section we ask: how does it affect the overall normalization 
of the likelihood? This is an inherently nonperturbative question. 

Since the galaxy field $\smash{\tilde{\delta}_g}$ is now filtered at the scale $\Lambda$, it is clear that the integral 
of the likelihood over it is not equal to $1$ unless the factor in front of the exponential is modified from its infinite-$\Lambda$ 
expression of Section~\ref{sec:main_result}. This is a problem because, as we discussed in 
Section~\ref{subsec:zero_noise_normalization_etc}, having the correct normalization ensures that we recover 
the correct limit of a Dirac delta functional when $\smash{P^{\{0\}}_{\eps_g}}$ goes to zero. Moreover, 
an incorrect normalization would lead to the wrong maximum-likelihood point for the parameters in the 
covariance (e.g.~$\smash{P^{\{0\}}_{\eps_g}}$ itself). 

We will now show that, if we include the full dependence of the 
covariance on the matter fields, the presence of a cutoff in Fourier space 
makes it difficult to normalize the likelihood with respect to the data 
while retaining a closed analytical form for it. Then, we discuss some 
ways in which one can get around this problem. 

Let us consider the exponential in \eqsII{intro}{normalization-5}, or in \eq{stochasticity_bias_coefficients-4} 
when we want to include the modulation of the noise by the matter field. As we discussed in Section~\ref{subsec:cutoffs} 
the galaxy field and its deterministic bias expansion are cut, and both the covariance and the deterministic bias 
expansion are constructed from the filtered matter field. However, as far as the integral over the galaxy field is concerned, 
the only important cuts are those on $\smash{\tilde{\delta}_g}$ and $\smash{\tilde{\delta}_{g,{\rm det}}}$. More 
precisely, in order to check the normalization we only need to check if the integral of the functional 
\begin{equation}
\label{eq:covariance-1}
\exp\Bigg({-\frac{1}{2}}\int\dif^3\tilde{x}\,\frac{\big(\chi_\Lambda(\tilde{\vec{x}}) 
- \varphi_\Lambda(\tilde{\vec{x}})\big)^2}{P(\tilde{\vec{x}})}\Bigg) 
\end{equation} 
over $\smash{\chi_\Lambda(\tilde{\vec{x}})}$ is well defined. 

First, we can perform a field redefinition $\smash{\chi'_\Lambda(\tilde{\vec{x}}) = \chi_\Lambda(\tilde{\vec{x}}) 
- \varphi_\Lambda(\tilde{\vec{x}})}$. Given that both fields are cut at $\Lambda$, the new field $\chi'$ also has no 
support for momenta below $\Lambda$, and the functional measure does not change. Dropping the subscript for simplicity of 
notation, we now have to integrate 
\begin{equation}
\label{eq:covariance-2}
\exp\Bigg({-\frac{1}{2}}\int\dif^3\tilde{x}\,\frac{\chi^2_\Lambda(\tilde{\vec{x}})}{P(\tilde{\vec{x}})}\Bigg) 
\end{equation} 
over $\smash{\chi_\Lambda(\tilde{\vec{x}})}$. Let us define 
\begin{equation}
\label{eq:covariance-3}
\frac{\chi_\Lambda(\tilde{\vec{x}})}{\sqrt{P(\tilde{\vec{x}})}} = \xi(\tilde{\vec{x}})\,\,, 
\end{equation} 
where we used the fact that $P(\tilde{\vec{x}})$ is positive, cf.~Section~\ref{subsec:perturbativity}. 
While a simple rescaling at the field level, this is a fully nonlinear redefinition in $\smash{\tilde{\vec{x}}}$. 
Therefore $\xi$ will contain all wavelengths even if the field $\chi_\Lambda$ is filtered. 
The Jacobian of the field redefinition does not depend on $\xi$, so the integral we are after is equal to 
\begin{equation}
\label{eq:covariance-4}
{\abs[\bigg]{\dfrac{\partial\chi_\Lambda(\tilde{\vec{x}})}{\partial\xi(\tilde{\vec{x}}')}}}\, 
\underbrace{\int{\cal D}\xi\,\exp\bigg({-\frac{1}{2}}\int\dif^3\tilde{x}\,\xi(\tilde{\vec{x}})\bigg)}_{
\hphantom{\prod_{\tilde{\vec{x}}}\sqrt{2\pi}\,}=\,\prod_{\tilde{\vec{x}}}\sqrt{2\pi}}\,\,, 
\end{equation} 
where $\smash{{\cal D}\xi = \prod_{\tilde{\vec{x}}}{\rm d}\xi(\tilde{\vec{x}})}$. 

What about the Jacobian of the field redefinition? In Appendix~\ref{app:functional_change_of_coordinates} we show that 
\begin{equation}
\label{eq:covariance-5}
\dfrac{\partial\chi_\Lambda(\tilde{\vec{x}})}{\partial\xi(\tilde{\vec{x}}')} = 
\sqrt{P(\tilde{\vec{x}}')}\,\widehat{W}_\Lambda\big(\abs{\tilde{\vec{x}}-\tilde{\vec{x}}'}\big)\,\,, 
\end{equation} 
where $\smash{\widehat{W}_\Lambda}$ is the Fourier transform of the filter $W_\Lambda(k)$ defined by 
\begin{equation}
\label{eq:covariance-6}
\chi_\Lambda(\tilde{\vec{x}}) = \int_{\vec{k}}W_\Lambda(k)\,\chi(\vec{k})\,\eu^{\iu\vec{k}\cdot\tilde{\vec{x}}}\,\,. 
\end{equation}
The field redefinition, and consequently the functional integral itself, is well defined 
only if this ``matrix'' is positive-definite (technically, if this is the kernel of a positive-definite 
linear operator in the space of square-integrable functions). Let us consider for example a filter 
$\smash{W_\Lambda(k) = W(k^2/\Lambda^2)}$. This encompasses the most common filters like a hard cut, 
as in \eq{like_review-3}, or a Gaussian filter. It is then easy to convince oneself that, for a generic 
$\smash{P(\tilde{\vec{x}}')}$, \eq{covariance-5} is not a positive-definite matrix. 
A quick way to see this is the following. In \eq{covariance-5} we know that, for $\smash{W_\Lambda(k) = W(k^2/\Lambda^2)}$, 
the Fourier transform of the filter can be written as a series expansion in $\nabla^2/\Lambda^2$, that is 
(see also Appendix~\ref{app:functional_change_of_coordinates}) 
\begin{equation}
\label{eq:covariance-footnote-1}
\widehat{W}_\Lambda\big(\abs{\tilde{\vec{x}}-\tilde{\vec{x}}'}\big) = 
\delta^{(3)}_{\rm D}(\tilde{\vec{x}}-\tilde{\vec{x}}') 
+ \sum_{n=1}^{+\infty}c_n\frac{(-1)^n}{\Lambda^{2n}} 
\nabla^{2n}\delta^{(3)}_{\rm D}(\tilde{\vec{x}}-\tilde{\vec{x}}')\,\,, 
\end{equation} 
where the $c_n$ are defined via 
\begin{equation}
\label{eq:covariance-footnote-2}
W\bigg(\frac{k^2}{\Lambda^2}\bigg) = 1+\sum_{n=1}^{+\infty} c_n \frac{k^{2n}}{\Lambda^{2n}}\,\,. 
\end{equation} 
If we use the properties of the Dirac delta to move the derivatives on $\smash{\sqrt{P(\tilde{\vec{x}}')}}$, 
we see that we end up with a diagonal matrix (which also implies that we can easily compute the Jacobian of 
the field redefinition using the property $\ln\det=\Tr\ln$). 
However, while $\smash{\sqrt{P(\tilde{\vec{x}}')}}$ is positive for all 
$\smash{\tilde{\vec{x}}'}$, there is no guarantee that this is true also 
after the infinite series of derivatives of \eq{covariance-footnote-1} acts on it. 

While we will investigate this in more detail in future work as well, 
it is important to emphasize that it is not a showstopper. Indeed, we 
can already identify two ways in which this problem can be addressed. 

First, it is what we learned about the perturbative scalings in the previous section 
(that mirrors what Refs.~\cite{Cabass:2019lqx,Cabass:2020nwf} also discuss) that comes 
to our rescue. We have seen that it is always more important to include nonlinearities 
in the forward model than nonlinearities in the field-dependent covariance. More 
precisely, \eq{perturbativity-4} tells us that second-order terms in 
$\smash{\tilde{\delta}_{g,{\rm det}}[\delta,\!\vec{v}](\tilde{\vec{x}})}$ are enhanced by 
\begin{equation}
\label{eq:covariance-7}
\bigg(\frac{1}{k_{\rm NL}P^{\{0\}}_{\eps_g}}\bigg)^{\frac{3}{2}}\bigg(\frac{\Lambda}{k_{\rm NL}}\bigg)^{\frac{n_\delta}{2}} 
\end{equation} 
with respect to linear terms in $\smash{\tilde{J}[\delta,\!\vec{v}](\tilde{\vec{x}})}$. 
Comparing with cubic terms in $\smash{\tilde{\delta}_{g,{\rm det}}[\delta,\!\vec{v}](\tilde{\vec{x}})}$, 
instead, would give an additional $\smash{\sim(\Lambda/k_{\rm NL})^{(3+n_\delta)/2}}$, leading to 
\begin{equation}
\label{eq:covariance-8}
\bigg(\frac{1}{k_{\rm NL}P^{\{0\}}_{\eps_g}}\bigg)^{\frac{3}{2}}\bigg(\frac{\Lambda}{k_{\rm NL}}\bigg)^{\frac{3}{2}+n_\delta}\,\,. 
\end{equation} 
This is close to being $\Lambda$-independent for ${-2}\lesssim n_\delta\lesssim{-1.5}$. 

Therefore we conclude that, as long as we stop at second order in perturbations in 
the deterministic bias expansion, we take $\smash{\tilde{J}[\delta,\!\vec{v}](\tilde{\vec{x}}) = 1}$ 
in \eqsII{intro}{normalization-5},\footnote{Equivalently, in \eq{stochasticity_bias_coefficients-4} we take 
$\smash{\tilde{P}_\eps[\delta,\!\vec{v}](\tilde{\vec{x}})/ 
\tilde{J}[\delta,\!\vec{v}](\tilde{\vec{x}})}$ equal to $\smash{P^{\{0\}}_{\eps_g}}$.} 
and we take $\Lambda$ such that the dimensionless number in \eq{covariance-7} is small, 
we are sure that we are not neglecting terms in the likelihood that are as (or more) important 
than the ones we are keeping. Since we are now effectively in the situation where 
$\smash{\sqrt{P(\tilde{\vec{x}}')}}$ is independent of $\smash{\tilde{\vec{x}}'}$, 
there are no problems with the normalization of the likelihood with respect to the data: it goes 
through straightforwardly as discussed in Section~\ref{subsec:eft_likelihood_review}. 

A second solution, that does not require stopping at a finite order in perturbations, 
is the following. Let us consider a cubic filter $\smash{W_\Lambda(\vec{k})}$, defined as 
\begin{equation}
\label{eq:covariance-9}
W_\Lambda(\vec{k}) = \prod_{i=1}^3\Theta_{\rm H}\big(\Lambda-\abs{k^i}\big)\,\,. 
\end{equation} 
The calculations in Appendix~\ref{app:functional_change_of_coordinates} 
go through in the same way, so that \eq{covariance-5} becomes 
\begin{equation}
\label{eq:covariance-10}
\dfrac{\partial\chi_\Lambda(\tilde{\vec{x}})}{\partial\xi(\tilde{\vec{x}}')} = 
\sqrt{P(\tilde{\vec{x}}')}\,\widehat{W}_\Lambda(\tilde{\vec{x}} - \tilde{\vec{x}}')\,\,, 
\end{equation} 
with 
\begin{equation}
\label{eq:covariance-11}
\widehat{W}_\Lambda(\tilde{\vec{x}} - \tilde{\vec{x}}') = 
\frac{1}{\pi^3}\prod_{i=1}^3\frac{\sin\big(\Lambda(\tilde{x}_i-\tilde{x}'_i)\big)}{\tilde{x}_i-\tilde{x}'_i}\,\,. 
\end{equation} 
Let us then put the fields on a lattice, $\smash{\tilde{\vec{x}}=a\vec{l}}$ and $\smash{\tilde{\vec{x}}'=a\vec{l}'}$ 
($\smash{\vec{l},\vec{l}'\in\mathbb{N}^3}$), with lattice spacing $\smash{a=2\pi/\Lambda}$. 
It is straightforward to see that the matrix of \eq{covariance-10} is 
now diagonal in $\smash{\vec{l},\vec{l}'}$: the filter is a positive number proportional to 
$\smash{\Lambda^3}$ for $\smash{\vec{l} = \vec{l}'}$, and it is equal to zero if $\smash{\vec{l}\neq\vec{l}'}$. 
Since $\smash{\sqrt{P(\tilde{\vec{x}}')}}$ remains positive also when evaluated on a lattice, 
\eq{covariance-10} is a positive-definite matrix and the functional integral giving the 
normalization of the likelihood is well defined.

%********************************************************
\section{Stochasticity in the galaxy velocity field}
\label{sec:velocity_noise}
%********************************************************

\noindent In this section we discuss what is the effect of the stochasticity in $\smash{\vec{v}_g}$. As we 
discussed in Section~\ref{subsec:notation_and_conventions}, this stochasticity is guaranteed to be subleading 
in derivatives by the equivalence principle (for more details we refer to \cite{Perko:2016puo} and Section~2.8 
of \cite{Desjacques:2016bnm}). Following the notation of \cite{Desjacques:2016bnm}, we write 
\begin{equation}
\label{eq:velocity_noise-1}
\vec{v}_g(\vec{x}) = \vec{v}_{g,{\rm det}}[\delta,\!\vec{v}](\vec{x}) + \vec{\varepsilon}_v(\vec{x})\,\,. 
\end{equation}
Then, the fact that the source of relative acceleration between galaxies and matter can only be a functional of 
$\smash{\vec{\nabla}\delta(\vec{x})}$ (and of derivatives of the tidal field at higher order in perturbations) 
guarantees that the power spectrum of $\smash{\vec{\varepsilon}_v}$ is of the form 
\begin{equation}
\label{eq:velocity_noise-2}
P_{\eps^i_v\eps^j_v}(k) = P^{\{2\}}_{0}k^ik^j + P^{\{2\}}_{\pm 1}(k^2\delta^{ij} - k^ik^j)\,\,, 
\end{equation} 
where the constants $\smash{P^{\{2\}}_{0}}$ and $\smash{P^{\{2\}}_{\pm 1}}$ have dimensions 
of a length to the $5$th power. Similarly, the cross-correlation with $\eps_g$ satisfies 
\begin{equation}
\label{eq:velocity_noise-3}
P_{\eps^i_v\eps_g}(k) = \iu k^iP^{\{1\}}_{\eps_v\eps_g}\,\,,
\end{equation} 
where $\smash{P^{\{1\}}_{\eps_v\eps_g}}$ has dimensions of a length to the $4$th power. 

Before proceeding, let us explain the choice of notation in \eq{velocity_noise-2}. First, we emphasize that 
the absence of preferred directions allows both a term $\smash{\propto k^2\delta^{ij}}$ and one 
$\smash{\propto k^ik^j}$.\footnote{Notice that neither \cite{Perko:2016puo} nor \cite{Desjacques:2016bnm} 
included the term $\smash{\propto k^2\delta^{ij}}$. This term has been discussed in detail in 
\cite{Mercolli:2013bsa}.} Let us decompose $\smash{\vec{\varepsilon}_v(\vec{x})}$ 
in a longitudinal part and a transverse part, i.e.~ 
\begin{equation}
\label{eq:velocity_noise-4}
\vec{\varepsilon}_v(\vec{x}) = \vec{\nabla}\eps_{0}(\vec{x}) + \vec{\nabla}\times\vec{\eps}_{\pm 1}(\vec{x})\,\,. 
\end{equation} 
Then, \eqsII{velocity_noise-2}{velocity_noise-3} are equivalent to having 
\begin{subequations}
\label{eq:velocity_noise-5}
\begin{align}
P_{\eps_{0}}(k) &= {-\smash{P^{\{2\}}_{0}}}\,\,, \label{eq:velocity_noise-5-1} \\
P_{\eps^i_{\pm 1}\eps^j_{\pm 1}}(k) &= {-\smash{P^{\{2\}}_{\pm 1}}}\delta^{ij}\,\,, \label{eq:velocity_noise-5-2} \\
P_{\eps^i_{\pm 1}\eps_{0}}(k) &= 0\,\,, \label{eq:velocity_noise-5-3} \\ 
P_{\eps_{0}\eps_g}(k) &= {P^{\{1\}}_{\eps_v\eps_g}}\,\,. \label{eq:velocity_noise-5-4} 
\end{align}
\end{subequations} 
That is, the term $\smash{\propto k^2\delta^{ij}}$ comes from the transverse part of the noise field 
$\smash{\vec{\varepsilon}_v(\vec{x})}$, where we used the fact that its constant power spectrum must 
be proportional to $\delta^{ij}$ and its correlation with $\smash{\eps_{0}(\vec{x})}$ must vanish 
because of the absence of preferred directions. 

We will see why this term is important in Section~\ref{subsec:impact_on_eft_likelihood} below. 
Before doing that, however, let us build some intuition for what the leading corrections to 
the EFT likelihood from \eqsII{velocity_noise-2}{velocity_noise-3} are. At the linear level in 
perturbations, from \eq{perturbativity-2} we have 
\begin{equation}
\label{eq:intuition-1}
\tilde{\delta}_{g,{\rm det}}(\tilde{\vec{x}}) = \delta_{g,{\rm det}}[\delta](\tilde{\vec{x}}) 
- \frac{\vers{n}\cdot\partial_\parallel 
\vec{v}_{g,{\rm det}}[\delta,\!\vec{v}](\tilde{\vec{x}})}{\cal H} 
+ \eps_g(\tilde{\vec{x}}) - \frac{\vers{n}\cdot\partial_\parallel\vec{\eps}_v(\tilde{\vec{x}})}{\cal H}\,\,. 
\end{equation} 
From this, we see that at this order the noise in $\smash{\tilde{\delta}_g}$ is 
\begin{equation}
\label{eq:intuition-2}
\eps_g(\tilde{\vec{x}}) - \frac{\vers{n}\cdot\partial_\parallel\vec{\eps}_v(\tilde{\vec{x}})}{\cal H}\,\,. 
\end{equation} 
Its power spectrum is a sum of three terms. The first is the rest-frame noise power 
spectrum. Then, we have the power spectrum of $\smash{\vec{\eps}_v}$, and finally its 
correlation with $\smash{\eps_g}$. Using \eqsIII{velocity_noise-2}{velocity_noise-3}{velocity_noise-5} 
we see that the contribution from the correlation of the rest-frame noise with the noise in $\smash{\vec{v}_g}$ 
is less suppressed in derivatives with respect to the contribution from the power spectrum of $\smash{\vec{\eps}_v}$. 
More precisely, its scaling with $k^2$ is the same as the leading higher-derivative corrections to the 
rest-frame noise power spectrum, cf.~\eqsII{like_review-1}{like_review-2} (notice that we have a power of 
$\smash{\vec{k}\cdot\vers{n} = k\mu}$ from $\smash{\partial_\parallel}$, and another power of $\smash{k\mu}$ 
from $\smash{\sum_i n^iP_{\eps^i_v\eps_g}(k)}$: hence the dependence is on $\smash{k^2\mu^2}$). 
The terms coming from $\smash{P_{\eps^i_v\eps^j_v}(k)}$ carry an additional $k^2$ suppression 
(and give a $\smash{\mu^2}$ or $\smash{\mu^4}$ angular dependence). 

In the next three sections we show how to derive these higher-derivative corrections 
to the likelihood. Moreover, we discuss what are the physical scales that suppress them.

%********************************************************
\subsection{Impact on the EFT likelihood}
\label{subsec:impact_on_eft_likelihood}
%********************************************************

\noindent Let us now study the impact of $\smash{\vec{\varepsilon}_v(\vec{x})}$ 
on the EFT likelihood in more detail. The manipulations in Section~\ref{sec:main_result} relied 
strongly on the the fact that the noise was local: since we are now dealing with 
higher-derivative corrections we cannot resum them at all orders in perturbations 
even if we work in the same infinite-$\Lambda$ limit of Section~\ref{sec:main_result}. 
For this reason we will only sketch how to derive the more straightforward of 
these corrections and, most importantly, we will show in the next section that they 
are under perturbative control. Moreover, we will mostly work in Fourier space 
throughout this section since this makes manipulating higher-derivative terms a much easier task. 

First, let us consider the noise correlators of \eqsII{velocity_noise-2}{velocity_noise-3}, 
together with the constant part of $\smash{P_{\eps_g}(k)}$ (we refer to \cite{Cabass:2020nwf} 
for a discussion on how to perturbatively include its higher-derivative corrections). 
If we construct the covariance matrix in Fourier space for $\smash{\eps_g}$ and 
$\smash{\vec{\varepsilon}_v}$ it is straightforward to see that its determinant is equal to 
\begin{equation}
\label{eq:velocity_noise-6}
{\Big(\smash{P^{\{2\}}_{\pm 1}}\Big)^2}\smash{P^{\{0\}}_{\eps_g}P^{\{2\}}_{0}}k^6 
+ {\Big(\smash{P^{\{2\}}_{\pm 1}}{P^{\{1\}}_{\eps_v\eps_g}}\Big)^2}k^6\,\,. 
\end{equation} 
That is, if $\smash{P^{\{2\}}_{\pm 1} = 0}$ we have gauge issues if we work with $\smash{\vec{\varepsilon}_v}$ 
as noise variable, and we must switch to a likelihood for $\smash{\eps_{0}}$ only. 

A simplification arises if we consider the case where the velocity noise is uncorrelated with $\eps_g$ 
and we take (introducing $P_v$ to simplify the otherwise very heavy notation) 
\begin{equation}
\label{eq:velocity_noise-7}
P^{\{2\}}_{\pm 1} = P^{\{2\}}_{0}\equiv P_v\,\,. 
\end{equation}
The likelihood $\smash{{\cal P}[\eps_g,\vec{\eps}_v]}$ then factorizes in 
\begin{equation}
\label{eq:velocity_noise-8}
{\cal P}[\eps_g,\vec{\eps}_v] = {\cal P}[\eps_g]{\cal P}[\vec{\eps}_v]\,\,. 
\end{equation} 
This leads to a great simplification. Indeed, we can work separately with $\smash{\eps_g}$ (whose likelihood we 
will still keep as in \eq{integrating_out_noise-2}, i.e.~in real space) and with $\smash{\vec{\eps}_v}$. 
The likelihood for the latter is given by 
\begin{equation}
\label{eq:velocity_noise-9}
{\cal P}[\vec{\eps}_v] = \Bigg(\prod_{\vec{k}}\frac{1}{(2\pi)^{3/2}P_v^3k^3}\Bigg) 
\exp\Bigg({-\frac{1}{2}}\int_{\vec{k}}\frac{\abs{\vec{\eps}_v(\vec{k})}^2}{P_vk^2}\Bigg)\,\,. 
\end{equation}
Let us see how this leads to a derivative expansion for the EFT likelihood. 
First, we have the equivalent of Section~\ref{subsec:integrating_out_noise}. 
Very schematically, we write this as 
\begin{equation}
\label{eq:velocity_noise-10}
{\cal P}[\tilde{\delta}_g|\delta,\!\vec{v}] = \int{\cal D}\eps_g{\cal D}\vec{\eps}_v\, 
{\cal P}[\eps_g,\vec{\eps}_v]\,\delta^{(\infty)}_{\rm D}\big(\tilde{\delta}_g 
- \tilde{\delta}_{g,{\rm det}} - {\rm noise}\big)\,\,. 
\end{equation} 
What is important in this equation is that the we take the argument of the Dirac delta functional to be exactly 
as in Section~\ref{subsec:integrating_out_noise}: the only difference is that now the galaxy velocity field 
is not equal to its deterministic bias expansion, but is instead given by \eq{velocity_noise-1}, i.e.~ 
\begin{equation}
\label{eq:velocity_noise-11}
\vec{v}_g(\vec{x}) = \vec{v}_{g,{\rm det}}[\delta,\!\vec{v}](\vec{x}) + \vec{\varepsilon}_v(\vec{x})\,\,. 
\end{equation} 
The integral over the noise $\smash{\eps_g}$ goes through in the same way. We then arrive at the equivalent 
of \eq{integrating_out_noise-8}, that is\footnote{For simplicity we are not going to include the stochasticity 
of bias coefficients. Our conclusions can be straightforwardly extended to account for a field-dependent noise 
covariance in the rest frame.} 
\begin{equation} 
\label{eq:velocity_noise-12} 
\Bigg(\prod_{\vec{x}}\sqrt{\frac{1}{2\pi P_{\eps_g}^{\{0\}}}}\,\Bigg) 
\exp\Bigg({-\frac{1}{2}}\int\dif^3x\,\frac{\big\{1 + 
\tilde{\delta}_g\big(\mathscr{R}^{-1}(\vec{x})\big) 
- \big(1+\delta_{g,{\rm det}}[\delta](\vec{x})\big)/J[\vec{v}_g](\vec{x})\big\}^2} 
{P_{\eps_g}^{\{0\}}/J^2[\vec{v}_g](\vec{x})}\Bigg)\,\,. 
\end{equation} 
Here we emphasized the dependence of the Jacobian $\smash{J}$ of the coordinate change on the full galaxy velocity field of 
\eq{velocity_noise-11} by changing its argument from $\smash{J[\delta,\!\vec{v}]}$ to $\smash{J[\vec{v}_g]}$. Also, at 
variance with \eq{notation-9-A}, in the above equation we have 
\begin{equation} 
\label{eq:velocity_noise-13} 
\mathscr{R}[\vec{v}_g]\equiv\mathscr{R}\,\,. 
\end{equation} 

It is then only a matter of carrying out the integral over $\smash{\vec{\eps}_v}$. We can do it perturbatively 
in derivatives by using a functional generalization of \eq{covariance-footnote-1}. In 
Appendix~\ref{app:tricks_for_velocity_noise-A} we show that 
\begin{equation}
\label{eq:velocity_noise-14}
{\cal P}[\vec{\eps}_v] = \exp\bigg({-\frac{1}{2}}\int_{\vec{k}}P_vk^2\,\frac{\partial}{\partial\vec{\eps}_v(\vec{k})} 
\cdot\frac{\partial}{\partial\vec{\eps}_v({-\vec{k}})}\bigg)\,\delta^{(\infty)}_{\rm D}\big(\vec{\eps}_v(\vec{k})\big)\,\,. 
\end{equation} 
When we integrate \eq{velocity_noise-12} against $\smash{{\cal P}[\vec{\eps}_v]}$ we can expand the exponential and 
use the properties of the Dirac delta functional to move the derivatives from 
$\smash{\delta^{(\infty)}_{\rm D}\big(\vec{\eps}_v(\vec{k})\big)}$ to \eq{velocity_noise-12}, 
arriving then at a series expansion for the conditional likelihood. 
Appendix~\ref{app:tricks_for_velocity_noise-B} shows how this works in practice. 
However, the results found so far are enough to discuss what are the parameters we are expanding the likelihood in: 
this is the subject of the next section.

%********************************************************
\subsection{Checking the perturbative expansion}
\label{subsec:checking_the_perturbative_expansion}
%********************************************************

\noindent As in Section~\ref{subsec:cutoffs} we reintroduce a filter at the scale $\Lambda$, 
applied to all the fields in \eq{velocity_noise-12} (including the noise $\smash{\vec{\eps}_v}$). 
It is then sufficient to study the scaling dimensions of the objects in the exponential of \eq{velocity_noise-14} 
under a rescaling $\vec{k}\to b\vec{k}$ (with $b\leq 1$) to find how the size of the 
corrections becomes smaller at decreasing $\Lambda$.\footnote{Since we are working perturbatively, 
there is no need to consider how loop corrections could affect the scaling dimensions 
(see also \cite{Pajer:2013jj} for details).} 
\begin{itemize}[leftmargin=*]
\item First, we have $\int_{\vec{k}}$. Under $\vec{k}\to b\vec{k}$, $\dif^3k$ scales as $b^3$. 
\item We then have an additional $b^2$ from $k^2$. 
\item Finally, we have the two functional derivatives. We can see that they are invariant under the rescaling 
$\vec{k}\to b\vec{k}$ thanks to the equivalent of \eq{notation-11}, i.e.~ 
\begin{equation}
\label{eq:velocity_noise-15}
\frac{\partial\chi(\vec{k})}{\partial\chi(\vec{k}')} = (2\pi)^3\delta^{(3)}_{\rm D}(\vec{k}+\vec{k}')\,\,. 
\end{equation} 
Given that both the Fourier modes $\smash{\chi(\vec{k})}$ of the dimensionless field $\smash{\chi(\vec{x})}$ 
and $\smash{\delta^{(3)}_{\rm D}(\vec{k}+\vec{k}')}$ have dimension of a length to $3$rd power, 
we see that $\smash{\partial/\partial\chi(\vec{k}')}$ is dimensionless. 
Then, the only question is what is the ``typical scale of variation'' of \eq{velocity_noise-12} with 
respect to the noise in the galaxy velocity field. From Section~\ref{subsec:rsds}, more precisely 
\eqsII{RS_review-1}{RS_review-2}, we see that every occurrence of the galaxy velocity field comes 
with an additional $\smash{\partial_\parallel/{\cal H}}$. Since we have two functional derivatives, we have 
an additional $\smash{b^2}$ scaling, controlled by $\smash{\mu^2/{\cal H}^2}$. It is important to emphasize 
that derivatives with respect to $\smash{\vec{\eps}_v}$ can lead to additional insertions of the fields 
$\smash{\delta}$, $\smash{\vec{\nabla}\vec{v}}$ or $\smash{\tilde{\delta}_g - \tilde{\delta}_{g,{\rm det}}}$. 
These contributions are less relevant in the infrared because they add powers of $\smash{\Lambda}$ according to 
\eqsIII{perturbativity-1}{perturbativity_check_scale}{perturbativity-3}: we will discuss them briefly in the next section. 
\end{itemize} 

What if we had not considered only the isotropic part ($\smash{\propto k^2\delta^{ij}}$) of 
$\smash{P_{\eps^i_v\eps^j_v}(k)}$? We would have obtained an additional overall $\smash{\mu^2}$. 
From this we conclude that the corrections to the likelihood from the 
power spectrum of the noise in the galaxy velocity field scale at least as 
\begin{equation}
\label{eq:velocity_noise-16}
\frac{\mu^2\Lambda^2}{{\cal H}^2}\,P_v\Lambda^5\,\,,\quad \frac{\mu^2\Lambda^2}{{\cal H}^2}\,P_v\mu^2\Lambda^5\,\,, 
\end{equation} 
as we had estimated via \eq{intuition-2}. We confirm this by explicit calculation in Appendix~\ref{app:tricks_for_velocity_noise-B}. 

The next question we have to ask, then, is what is the value we expect for $P_v$. 
A rough upper limit comes with the following argument (see Sections~2.7 and 2.8 of \cite{Desjacques:2016bnm} for more details). 
In the deterministic bias expansion we have $\smash{\vec{v}_g = \vec{v}}$ 
at linear order in perturbations and derivatives. This, in turn, is proportional to 
$\smash{({\cal H}\vec{\nabla}/\nabla^2)\delta}$. $\smash{\delta}$ is proportional, via $b_1$, to $\smash{\delta_g}$ at this order. 
Let us consider then the stochasticity of $\smash{\delta_g}$. At second order in derivatives it is $\smash{R_\ast^2\nabla^2\eps_g}$, 
so plugging this into $\smash{({\cal H}\vec{\nabla}/\nabla^2)\delta\sim({\cal H}\vec{\nabla}/\nabla^2)\delta_g}$ we get 
$\smash{\vec{\eps}_v\sim {\cal H}R^2_\ast\vec{\nabla}\eps_g}$, which means\footnote{A sharper estimate can be found 
by assuming a concrete physical model, e.g.~that the stochastic velocity contribution of a given galaxy sample 
is due to the virial velocities within the host halos of mass $M_h$ and Eulerian radius $R_E$ of these galaxies. 
This gives $P_v$ of order $R^5_E$ (that scales as $\smash{M_h^{5/3}}$). See e.g.~\cite{Schmidt:2010jr}, 
Section~2.8 of \cite{Desjacques:2016bnm} and Section~6 of \cite{Desjacques:2018pfv} for details.} 
\begin{equation}
\label{eq:velocity_noise-17} 
P_v\sim{\cal H}^2R^4_\ast P^{\{0\}}_{\eps_g}\,\,. 
\end{equation} 

In the next section we confirm that the corrections from the correlation of the 
velocity noise with $\smash{\eps_g}$ are more relevant on large scales than the ones 
of \eq{velocity_noise-16}. Moreover, we quickly discuss how to estimate the size of 
subleading terms coming from higher orders in $\smash{\vec{\eps}_v}$.

%********************************************************
\subsection{Leading correction from the noise in \texorpdfstring{$\vec{v}_g$}{v\_g}}
\label{subsec:leading_correction_from_velocity_noise}
%********************************************************

\noindent Let us discuss the impact of $\smash{\vec{\eps}_v}$ in the field-level 
relation between $\smash{\tilde{\delta}_g}$ and $\smash{\delta_g}$ of \eq{notation-3-B}, i.e.~ 
\begin{equation}
\label{eq:velocity_noise-18}
\tilde{\delta}_g(\tilde{\vec{x}}) = \frac{1+\delta_g\big(\mathscr{R}[\vec{v}_g](\tilde{\vec{x}})\big)}
{\tilde{J}[\vec{v}_g](\tilde{\vec{x}})} - 1\,\,. 
\end{equation} 
Writing the rest-frame galaxy density field as the sum of the deterministic bias 
expansion and the stochasticity $\smash{\eps_g}$, we see that the impact of $\smash{\vec{\eps}_v}$ 
is essentially twofold. Expanding $\smash{\vec{v}_g}$ around its deterministic bias expansion, 
we see that $\smash{\vec{\eps}_v}$ can either go to multiply $\smash{\delta_{g,{\rm det}}[\delta]}$ or 
$\smash{\eps_g}$, or can appear by itself (The Jacobian of the coordinate change in \eq{velocity_noise-18} 
is the most straightforward example). 
\begin{itemize}[leftmargin=*] 
\item If it multiplies the deterministic bias expansion for $\smash{\delta_g}$, its 
effect is basically the same as that of the stochasticity of the bias coefficients: 
it gives a modulation of the covariance suppressed by derivatives. 
\item If, on the other hand, $\smash{\vec{\eps}_v}$ multiplies $\smash{\eps_g}$ we 
have a contribution that is similar to that coming from the non-Gaussianity of $\smash{\eps_g}$ 
(again suppressed by derivatives). 
\item Then, the leading contribution is when $\smash{\vec{\eps}_v}$ appears by itself, e.g.~thanks 
to the fact that in \eq{velocity_noise-18} the Jacobian of the coordinate change to redshift space 
multiplies $\smash{1+\delta_g}$, and not only $\smash{\delta_g}$. This is exactly what gives the 
contribution shown in \eqsII{intuition-1}{intuition-2} (see also 
Appendix~\ref{app:tricks_for_velocity_noise-B} for more details). 
\end{itemize} 

Let us now see how to account for the fact that the rest-frame noise $\smash{\eps_g}$ and $\smash{\vec{\eps}_v}$ can be correlated, with 
\begin{equation}
\label{eq:velocity_noise-19}
P^{\{1\}}_{\eps_v\eps_g}\sim{\cal H}R_\ast^2P^{\{0\}}_{\eps_g} 
\end{equation}
via the same estimates that lead to \eq{velocity_noise-17}. 
The manipulations with functional integrals of Section~\ref{subsec:checking_the_perturbative_expansion} 
continue to hold, with the only differences being that in \eq{velocity_noise-14} $\smash{P_v k^2}$ is 
replaced by $\smash{P^{\{1\}}_{\eps_v\eps_g}\vec{k}\cdot\vers{n} = P^{\{1\}}_{\eps_v\eps_g}k\mu}$, 
and one derivative with respect to $\smash{\vec{\eps}_v}$ is replaced by a derivative 
with respect to $\smash{\eps_g}$. Consequently, we have one less power of $\mu\Lambda/{\cal H}$. 
As far as the scaling with $\Lambda$ is concerned, then, the scalings of \eq{velocity_noise-16} become 
\begin{equation}
\label{eq:velocity_noise-20}
\frac{\mu\Lambda}{{\cal H}}\,P^{\{1\}}_{\eps_v\eps_g}\mu\Lambda^4\,\,. 
\end{equation} 

An interesting point is comparing the scalings of \eqsII{velocity_noise-16}{velocity_noise-20} to the one from 
the higher-derivative corrections to the rest-frame noise power spectrum. The leading corrections there scale as 
$\smash{R_\ast^2\Lambda^2P^{\{0\}}_{\eps_g}\Lambda^3}$. Hence, we see from \eq{velocity_noise-17} that the ratio 
with the contribution from the power spectrum of $\smash{\vec{\eps}_v}$ scales as 
\begin{equation} 
\label{eq:velocity_noise-21} 
\frac{P_v\mu^2\Lambda^7}{{\cal H}^2R_\ast^2\Lambda^2P^{\{0\}}_{\eps_g}\Lambda^3}\sim R_\ast^2\mu^2\Lambda^2\,\,,\quad 
\frac{P_v\mu^4\Lambda^7}{{\cal H}^2R_\ast^2\Lambda^2P^{\{0\}}_{\eps_g}\Lambda^3}\sim R_\ast^2\mu^4\Lambda^2\,\,, 
\end{equation} 
which is always small for small $\smash{\Lambda}$ thanks to it 
being softer that the nonlocality scale of galaxy dynamics. 
As far as the contribution from the correlation of $\smash{\vec{\eps}_v}$ with $\smash{\eps_g}$ is concerned, instead, we have 
\begin{equation} 
\label{eq:velocity_noise-22} 
\frac{P^{\{1\}}_{\eps_v\eps_g}\mu^2\Lambda^5}{{\cal H}R_\ast^2\Lambda^2P^{\{0\}}_{\eps_g}\Lambda^3}\sim\mu^2\,\,, 
\end{equation} 
where we have used \eq{velocity_noise-19}. 

\begin{figure} 
\centering
\includegraphics[width=\columnwidth]{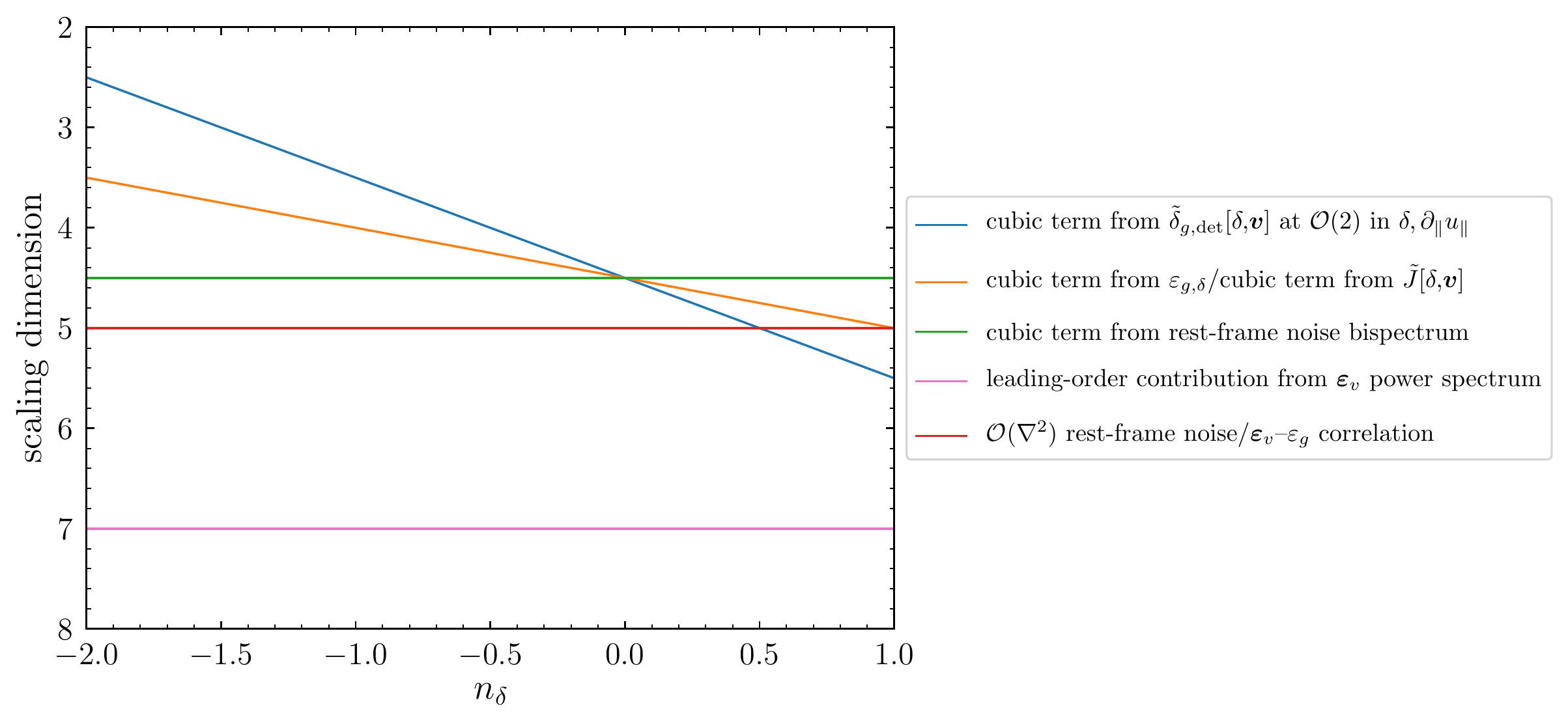} 
\caption{Scaling dimensions of the various contributions to $\smash{\ln{\cal P}[\tilde{\delta}_g|\delta,\!\vec{v}]}$. 
We plot them as a function of the spectral index $n_\delta$ of the linear power spectrum, ranging from $-2$, the 
approximate value of $n_\delta$ on very small scales, to $1$, the value of $n_\delta$ on scales much larger 
than the equality scale. The contributions considered here are at most of cubic order in 
the fields $\delta$, $\smash{\partial_\parallel u_\parallel}$ and $\smash{\tilde{\delta}_g(\tilde{\vec{x}}) 
- \big(b_1\delta(\tilde{\vec{x}}) - \partial_\parallel u_\parallel(\tilde{\vec{x}})\big)}$ (recall that 
$\smash{\tilde{\delta}_{g,{\rm det}}[\delta,\!\vec{v}](\tilde{\vec{x}}) = b_1\delta(\tilde{\vec{x}}) 
- \partial_\parallel u_\parallel(\tilde{\vec{x}})}$ at leading order in perturbations 
and derivatives, cf.~\eq{perturbativity-2}, and that \eqsII{perturbativity-1}{perturbativity_check_scale} tell us 
that $\smash{\delta(\tilde{\vec{x}})}$ and $\smash{\partial_\parallel u_\parallel(\tilde{\vec{x}})}$ scale 
in the same way with $\Lambda$). At this order we see, for example, that the stochasticity in 
the linear bias and the Jacobian of the coordinate change to redshift space are always less relevant than the deterministic 
evolution in the rest frame and the coordinate change itself. 
On the other hand, these are always more relevant than the non-Gaussianity of the rest-frame noise if $n_\delta$ is negative. 
In turn, we see that the noise in the galaxy velocity field itself is always subleading with respect to the 
non-Gaussianity of $\smash{\eps_g}$, and also with respect to the higher-derivative 
contributions to the rest-frame Gaussian noise (that scale in the same way as 
the contributions from a nonzero correlation between $\smash{\vec{\eps}_v}$ and $\smash{\eps_g}$).} 
\label{fig:scalings} 
\end{figure}

We conclude this Section with Fig.~\ref{fig:scalings}, that compares these scaling dimensions 
with those discussed in Section~\ref{subsec:perturbativity}.

%********************************************************
\section{Conclusions}
\label{sec:conclusions}
%********************************************************

\noindent In this paper we have studied how to implement redshift-space distortions 
in the EFT likelihood for large-scale structure. 
The equivalence principle forbids large-scale stochasticity in the galaxy velocity. 
Therefore, at leading order in a derivative expansion, the form of the likelihood 
is determined by the noise in the rest-frame galaxy density. The likelihood is still 
a Gaussian in the redshift-space galaxy density field $\smash{\tilde{\delta}_g(\tilde{\vec{x}})}$: 
the effect of redshift-space distortions is that of modifying its covariance. The 
galaxy noise power spectrum $\smash{P_{\eps_g}^{\{0\}}}$ is replaced by 
the field-dependent $\smash{P_{\eps_g}^{\{0\}}/\tilde{J}[\delta,\!\vec{v}](\tilde{\vec{x}})}$, 
where $\smash{\tilde{J}[\delta,\!\vec{v}](\tilde{\vec{x}})}$ is the Jacobian of the coordinate change 
$\smash{\vec{x}=\vec{x}(\tilde{\vec{x}})}$ to redshift space (which depends on the matter density and 
velocity fields through the bias expansion for the galaxy velocity). 

In the framework of the EFT-based forward modeling one wants the galaxy and matter fields 
to be cutoff at a scale $\Lambda$ (this is essentially what allows to make contact 
with the perturbative approach of the EFTofLSS), but also to sample 
the likelihood via, e.g., Hamiltonian Monte Carlo methods. 
This requires the likelihood to be normalized with respect to the data, i.e.~the 
redshift-space galaxy density field $\smash{\tilde{\delta}_g(\tilde{\vec{x}})}$. 
As long as we restrict our bias expansion and the coordinate change to redshift space to second order 
in perturbations, we show that it is consistent to neglect the dependence of the 
covariance of the Gaussian likelihood on the matter fields. The normalization of 
the likelihood can then be obtained in a closed form even when dealing with a filtered galaxy field. 
Putting the theory on a lattice with spacing $\smash{a=2\pi/\Lambda}$ is a second way to achieve this, 
its advantage being that it does not require to stop at a finite order in perturbation theory. 
This latter approach was implemented numerically (albeit for a rest-frame halo samples identified 
in N-body simulations) in the recent paper \cite{Schmidt:2020tao}. This paper checked the consistency between 
the real-space formulation of the likelihood and its Fourier-space counterpart of previous papers in this series, 
confirming that the two yield the same results for, e.g., the constraints on the amplitude of the primordial power 
spectrum (at fixed cosmological parameters and phases of the initial conditions). 

Finally, we have explicitly computed the corrections from the noise $\smash{\vec{\eps}_v}$ in the galaxy 
velocity field, confirming they are indeed subleading on large scales. Interestingly, we have shown that 
they are as relevant as the contribution of higher-derivative terms in the rest-frame noise 
power spectrum if $\smash{\vec{\eps}_v}$ is correlated with the noise in the rest-frame galaxy density. 

Two subjects that can be investigated in more detail in future work are selection effects and 
how to go beyond the flat-sky approximation (along with the interplay with the survey window function). 
However, we can already comment briefly about them. 
\begin{itemize}[leftmargin=*] 
\item The presence of the line-of-sight vector can alter the bias expansion $\smash{\delta_{g,{\rm det}}[\delta]}$ 
by adding a preferred direction to contract tensor indices with. For example, we can have 
the operator $\smash{\hat{n}^i\hat{n}^j K_{ij}[\delta]}$ appearing at linear order in perturbations. 
These terms arise when the selection function depends on the orientation of the galaxy 
\cite{Catelan:2000vm,Hirata:2007np,Okumura:2008du,Hirata:2009qz,Krause:2010tt,Fang:2011hc,Singh:2014kla}, 
or when galaxies are identified through emission or absorption lines, whose observed strength depends 
on the line of sight due to radiative transfer effects \cite{Zheng:2010jf,Wyithe:2011mt,Behrens:2017xmm}. 
This is not a problem, since the scaling dimensions for these operators are the same of those in 
Section~\ref{subsec:perturbativity}: it is then sufficient to identify all the operators at a given 
order in perturbations and include them into $\smash{\delta_{g,{\rm det}}[\delta]}$ 
(see e.g.~\cite{Desjacques:2018pfv}, and \cite{Agarwal:2020lov} for a recent study of their impact 
on constraints on cosmological parameters within the framework of the EFTofLSS). 
\item Finally, let us comment on the distant-observer approximation and the survey window function. 
First, none of our nonperturbative expressions for the likelihood of Section~\ref{sec:main_result} 
relied on the distant-observer approximation. Moreover, and most importantly, the scaling 
dimensions discussed in Sections~\ref{sec:cutoffs_and_perturbativity} and \ref{sec:velocity_noise} 
are also unaffected by it. Hence, we can understand the line of sight vector 
to be position-dependent throughout all of our expressions. A survey will in general probe a finite part of our past 
light cone, i.e.~a finite region in $\smash{(z,\vers{n})}$ space (for example, future large-scale galaxy surveys 
such as SPHEREx \cite{Dore:2014cca}, DESI \cite{Levi:2013gra}, and Euclid \cite{Amendola:2012ys,Amendola:2016saw} 
will have footprints of order $10000$ square degrees). Once we convert this region to its equivalent 
in $\smash{\tilde{\vec{x}}}$ coordinates, we can then simply integrate our likelihood over this region only, 
instead of over all $\smash{\tilde{\vec{x}}}$. 
\end{itemize} 
We can see, however, how this last point raises a very important issue, linked to the 
discussion of Section~\ref{subsec:covariance}. If we integrate our likelihood only over 
a subset of $\smash{\tilde{\vec{x}}}$ values, this is essentially equivalent to putting 
the noise power spectrum to infinity away from such region. That is, we now have effectively 
ended up with a position-dependent noise power spectrum. This seems to lead to the same complications 
that we encountered when discussing the dependence of the covariance on the matter density and 
velocity fields. The difference is that there is no expansion parameter that allows us to expand around an 
$\smash{\tilde{\vec{x}}}$-independent survey window function. Interestingly, since the window function 
is local in real space, it would not affect the conclusions of Section~\ref{subsec:covariance}, where we 
showed that the covariance matrix is diagonal if we put the theory on a lattice with spacing equal to $2\pi/\Lambda$. 

While we leave a more careful investigation to future work, it is important 
to emphasize that the complications discussed here arise because of the necessity 
of implementing the cutoff $\Lambda$ if one wants to make contact with EFT-based 
techniques. Bayesian forward modeling is already being applied to data from real 
surveys (e.g.~the SDSS-III/BOSS galaxy sample), therefore accounting for selection 
effects and the survey window function. For example, we refer to \cite{Jasche:2018oym,Ramanah:2018eed,Lavaux:2019fjr}, 
in which the forward model for redshift-space distortions is the same discussed 
here (see e.g.~Eq.~(18) of \cite{Jasche:2018oym}), and the differences with this 
work are essentially the likelihood for the galaxy distribution, the bias model, 
and especially the absence of the EFT cutoff $\Lambda$. The recent implementation 
of the EFT likelihood in real space of \cite{Schmidt:2020tao} suggests that it is 
feasible to introduce these ingredients in the analysis of \cite{Jasche:2018oym,Ramanah:2018eed,Lavaux:2019fjr}. 

We conclude with a comparison with the velocity reconstruction of \cite{Fisher:1994cm,Heavens:1994iq}. 
This reconstruction exploits the degree of anisotropy in a redshift-space galaxy catalog to measure the 
distortion from the expected real-space isotropy. One can reconstruct the matter field essentially by 
undoing the effect of the shift of \eq{RS_review-1} if linear theory, the continuity equation for 
matter and linear local-in-matter-density bias are assumed. From there, it is possible to measure, 
for example, the parameter $\smash{f/b_1}$ (where $f$ is the linear growth rate $\smash{f(a) = {\rm d}\ln D_1(a)/\dif\ln a}$). 
While \cite{Fisher:1994cm} works in real space and \cite{Heavens:1994iq} works in Fourier space 
(the latter having the advantage that, in linear theory, it is easy to restrict to long-wavelength, perturbative 
modes), both avoid the flat-sky approximation. In the language of Section~\ref{sec:notation_and_review} this 
amounts to considering $\smash{\vers{n}=\vers{n}(\vec{x})}$ (as discussed in the previous paragraph, 
none of the results in this paper depend on the flat-sky approximation). However, both works employ an expansion 
in spherical harmonics instead of using cartesian coordinates. While for a full-sky survey this is surely advantageous, 
it is less so once the partial sky coverage is taken into account. The formulation in cartesian coordinates of 
this paper and \cite{Schmidt:2020tao} should make the implementation of the window function more straightforward 
(see e.g.~Eq.~(2.6) of \cite{Schmidt:2020tao} for more details). 

Another difference between with \cite{Fisher:1994cm,Heavens:1994iq} and the approach discussed in this work 
lies in the fact that in the (EFT-based) forward modeling we do not move the data back to the rest-frame but 
we ``forward-model'' the rest-frame galaxy density to redshift space, and compare with data there. 
Finally, the EFT-based likelihood is constructed to go beyond linear perturbation theory, 
while still keeping nonlinear effects under control, thanks to the cutoff discussed in Section~\ref{sec:cutoffs_and_perturbativity}.

%********************************************************
\section*{Acknowledgements}
%********************************************************

\noindent It is a pleasure to thank Alex Barreira and Fabian Schmidt for very useful discussions, and Fabian Schmidt 
for collaboration on related topics. We acknowledge support from the Starting Grant (ERC-2015-STG 678652) ``GrInflaGal'' 
from the European Research Council.

%********************************************************
\appendix
%********************************************************

%********************************************************
\section{Integrating out the noise in redshift space}
\label{app:integrating_out_noise_appendix}
%********************************************************

\noindent In this appendix we derive again \eq{coordinate_change-10} by integrating out the noise, 
the difference with the calculation of Section~\ref{subsec:integrating_out_noise} being that we work 
directly in redshift space. Let us consider the relation 
\begin{equation}
\label{eq:integrating_out_noise_appendix-1}
\delta^{(\infty)}_{\rm D}\big(\tilde{\delta}_g(\tilde{\vec{x}}) 
- \tilde{\delta}_{g,{\rm det}}[\delta,\!\vec{v}](\tilde{\vec{x}}) - 
\tilde{\eps}_g(\tilde{\vec{x}})\big)\,\,, 
\end{equation} 
where 
\begin{equation}
\label{eq:integrating_out_noise_appendix-2}
\tilde{\eps}_g(\tilde{\vec{x}}) = \frac{\eps_g\big(\mathscr{R}(\tilde{\vec{x}})\big)} 
{\tilde{J}[\delta,\!\vec{v}](\tilde{\vec{x}})}\,\,. 
\end{equation}
We want to integrate \eq{integrating_out_noise_appendix-1} against the likelihood 
for $\smash{\tilde{\eps}_g(\tilde{\vec{x}})}$. Given \eq{integrating_out_noise_appendix-2} 
and the transformation rules for probability density functionals, the likelihood for the noise in redshift space is given by 
\begin{equation}
\label{eq:integrating_out_noise_appendix-3}
\abs[\bigg]{\frac{\partial\eps_g(\vec{x})}{\partial\tilde{\eps}_g(\tilde{\vec{x}}')}} 
\Bigg(\prod_{\vec{x}}\sqrt{\frac{1}{2\pi P_{\eps_g}^{\{0\}}}}\,\Bigg) 
\exp\Bigg({-\frac{1}{2}}\int\dif^3x\,\frac{\tilde{\eps}^2_g\big(\mathscr{R}^{-1}(\vec{x})\big)} 
{P^{\{0\}}_{\eps_g}/J^2[\delta,\!\vec{v}](\vec{x})}\Bigg)\,\,, 
\end{equation} 
where we have used the likelihood for the rest-frame noise of \eq{integrating_out_noise-2} and the 
inverse of \eq{integrating_out_noise_appendix-2} which, thanks to the definitions of 
\eqsII{notation-2}{notation-3-A}, is given by 
\begin{equation}
\label{eq:integrating_out_noise_appendix-4}
\eps_g(\vec{x}) = J[\delta,\!\vec{v}](\vec{x})\,\tilde{\eps}_g\big(\mathscr{R}^{-1}(\vec{x})\big)\,\,. 
\end{equation}

\eq{integrating_out_noise_appendix-4} straightforwardly allows us to compute the Jacobian in \eq{integrating_out_noise_appendix-3}. 
Using the definition of functional derivative, \eq{notation-11}, we get 
\begin{equation}
\label{eq:integrating_out_noise_appendix-5}
J[\delta,\!\vec{v}](\vec{x})\,\delta^{(3)}_{\rm D}(\mathscr{R}^{-1}(\vec{x})-\tilde{\vec{x}}') 
= \delta^{(3)}_{\rm D}(\vec{x}-\vec{x}')\,\,, 
\end{equation} 
where we used \eqsIII{coordinate_change-5}{coordinate_change-6}{coordinate_change-7}. The overall Jacobian is then equal to one. 
Then, in the integral of \eq{integrating_out_noise_appendix-3} we can change the argument to $\tilde{\vec{x}}$, 
and integrate in $\smash{{\cal D}\tilde{\eps}_g = \prod_{\tilde{\vec{x}}}\dif\tilde{\eps}_g(\tilde{\vec{x}})}$ 
against the Dirac delta functional of \eq{integrating_out_noise_appendix-1}. This gives \eq{coordinate_change-10} again.

%********************************************************
\section{Functional Jacobian matrix for a filtered field}
\label{app:functional_change_of_coordinates}
%********************************************************

\noindent In this short appendix we show how to arrive at \eq{covariance-5}. The field 
redefinition we have performed is the one of \eq{covariance-3}, which we can rewrite as 
\begin{equation}
\label{eq:app_functional_change_of_coordinates-1}
\chi_\Lambda(\tilde{\vec{x}}) = \Big(\!\sqrt{P(\tilde{\vec{x}})}\,\xi(\tilde{\vec{x}})\Big)_\Lambda\,\,, 
\end{equation} 
i.e.~ 
\begin{equation}
\label{eq:app_functional_change_of_coordinates-2}
\chi_\Lambda(\tilde{\vec{x}}) = 
\int_{\vec{k}}W_\Lambda(k)\,\bigg(\int_{\vec{q}}\xi(\vec{q})\mathscr{P}(\vec{k}-\vec{q})\bigg)\,
\eu^{\iu\vec{k}\cdot\tilde{\vec{x}}}\,\,, 
\end{equation} 
where we defined 
\begin{equation}
\label{eq:app_functional_change_of_coordinates-3}
\sqrt{P(\tilde{\vec{x}})} = \int_{\vec{k}}\mathscr{P}(\vec{k})\,\eu^{\iu\vec{k}\cdot\tilde{\vec{x}}} 
\end{equation}
and the filter $\smash{W_\Lambda(k)}$ is defined as 
\begin{equation}
\label{eq:app_functional_change_of_coordinates-4}
\chi_\Lambda(\tilde{\vec{x}}) = \int_{\vec{k}}W_\Lambda(k)\,\chi(\vec{k})\,\eu^{\iu\vec{k}\cdot\tilde{\vec{x}}}\,\,. 
\end{equation} 
Using the two relations 
\begin{subequations}
\label{eq:app_functional_change_of_coordinates-5}
\begin{align}
\frac{\partial\xi(\tilde{\vec{x}})}{\partial\xi(\tilde{\vec{x}}')} 
&= \delta^{(3)}_{\rm D}(\tilde{\vec{x}}-\tilde{\vec{x}}')\,\,, \label{eq:app_functional_change_of_coordinates-5-1} \\
\xi(\vec{q}) &= \int\dif^3\tilde{x}\,\xi(\tilde{\vec{x}})\, 
\eu^{-\iu\vec{q}\cdot\tilde{\vec{x}}}\,\,, \label{eq:app_functional_change_of_coordinates-5-2}
\end{align}
\end{subequations}
we have 
\begin{equation}
\label{eq:app_functional_change_of_coordinates-6}
\frac{\partial\xi(\tilde{\vec{q}})}{\partial\xi(\tilde{\vec{x}}')} = \eu^{-\iu\vec{q}\cdot\tilde{\vec{x}}'}\,\,. 
\end{equation} 
Hence, from \eq{app_functional_change_of_coordinates-2} we find 
\begin{equation}
\label{eq:app_functional_change_of_coordinates-7}
\begin{split}
\dfrac{\partial\chi_\Lambda(\tilde{\vec{x}})}{\partial\xi(\tilde{\vec{x}}')} &= 
\int_{\vec{k}}W_\Lambda(k)\,\eu^{\iu\vec{k}\cdot\tilde{\vec{x}}} 
\int_{\vec{q}}\eu^{-\iu\vec{q}\cdot\tilde{\vec{x}}'}\,\mathscr{P}(\vec{k}-\vec{q}) \\
&= \int_{\vec{k}}W_\Lambda(k)\,\eu^{\iu\vec{k}\cdot(\tilde{\vec{x}}-\tilde{\vec{x}}')}
\int_{\vec{p}}\mathscr{P}(\vec{p})\,\eu^{\iu\vec{p}\cdot\tilde{\vec{x}}'} \\
&= \widehat{W}_\Lambda\big(\abs{\tilde{\vec{x}}-\tilde{\vec{x}}'}\big)\, 
\sqrt{P(\tilde{\vec{x}}')}\,\,, 
\end{split} 
\end{equation} 
where in the last step we have used \eq{app_functional_change_of_coordinates-3}, 
and also that the filter is only a function of $\smash{k=\abs{\vec{k}}}$. 

As a byproduct of this derivation we can also obtain the series in \eq{covariance-footnote-1}. 
Indeed, if we expand the filter $W_\Lambda(k) = W(k^2/\Lambda^2)$ as 
\begin{equation}
\label{eq:app_functional_change_of_coordinates-8}
1+\sum_{n=1}^{+\infty} c_n \frac{k^{2n}}{\Lambda^{2n}}\,\,, 
\end{equation} 
we can write the first integral in the second-to-last line of \eq{app_functional_change_of_coordinates-7} as 
\begin{equation}
\label{eq:app_functional_change_of_coordinates-9}
\int_{\vec{k}}W_\Lambda(k)\,\eu^{\iu\vec{k}\cdot(\tilde{\vec{x}}-\tilde{\vec{x}}')} = 
\int_{\vec{k}}\bigg(1+\sum_{n=1}^{+\infty} c_n \frac{(-1)^n}{\Lambda^{2n}}\nabla^{2n}\bigg) 
\eu^{\iu\vec{k}\cdot(\tilde{\vec{x}}-\tilde{\vec{x}}')}\,\,,
\end{equation}
where the derivative can equivalently be taken with respect to $\smash{\tilde{\vec{x}}}$ 
or $\smash{\tilde{\vec{x}}-\tilde{\vec{x}}'}$ (or, using the fact that the filter 
is only a function of $\smash{k=\abs{\vec{k}}}$, with respect to $\vec{x}'$ or $\vec{x}'-\vec{x}$). 
Formally pulling the sum out of the integral sign, we get 
\begin{equation}
\label{eq:app_functional_change_of_coordinates-10}
\bigg(1+\sum_{n=1}^{+\infty} c_n \frac{(-1)^n}{\Lambda^{2n}}\nabla^{2n}\bigg) 
\delta^{(3)}_{\rm D}(\tilde{\vec{x}}-\tilde{\vec{x}}')\,\,. 
\end{equation}

%********************************************************
\section{Integrating out the velocity noise}
\label{app:velocity_noise}
%********************************************************

\noindent In this appendix we derive \eq{velocity_noise-14} and use it to compute one of the corrections 
to the conditional EFT likelihood $\smash{{\cal P}[\tilde{\delta}_g|\delta,\!\vec{v}]}$ 
by integrating out the velocity noise (Appendices~\ref{app:tricks_for_velocity_noise-A} and 
\ref{app:tricks_for_velocity_noise-B}, respectively).

%********************************************************
\subsection{Formal power series for the likelihood of the velocity noise}
\label{app:tricks_for_velocity_noise-A}
%********************************************************

\noindent Let us consider \eq{velocity_noise-9}, i.e.~ 
\begin{equation}
\label{eq:tricks_for_velocity_noise-A-1}
{\cal P}[\vec{\eps}_v] = \Bigg(\prod_{\vec{k}}\frac{1}{(2\pi)^{3/2}P_v^3k^3}\Bigg) 
\exp\Bigg({-\frac{1}{2}}\int_{\vec{k}}\frac{\abs{\vec{\eps}_v(\vec{k})}^2}{P_vk^2}\Bigg)\,\,. 
\end{equation} 
We can define its functional Fourier transform by 
\begin{equation}
\label{eq:tricks_for_velocity_noise-A-2}
{\cal P}[\vec{\eps}_v] = \int{\cal D}\mathbf{E}\,\Bigg(\prod_{\vec{k}}\frac{1}{(2\pi)^3}\Bigg)\, 
{\cal P}[\mathbf{E}]\,\exp\bigg(\iu\int_{\vec{k}}\mathbf{E}(\vec{k})\cdot\vec{\eps}_v({-\vec{k}})\bigg)\,\,,
\end{equation}
where $\smash{{\cal D}\mathbf{E}=\prod_{\vec{k}}\dif\mathbf{E}(\vec{k})}$ and 
we notice that $\smash{\mathbf{E}(\vec{k})}$ is dimensionless. 
Since the likelihood $\smash{{\cal P}[\vec{\eps}_v]}$ is a normalized Gaussian, 
we know that its functional Fourier transform must take the form 
\begin{equation}
\label{eq:tricks_for_velocity_noise-A-3}
{\cal P}[\mathbf{E}] = \exp\Bigg({-\frac{1}{2}}\int_{\vec{k}}P_vk^2\abs{\mathbf{E}(\vec{k})}^2\Bigg)\,\,,
\end{equation}
i.e.~it is equal to $1$ for vanishing $\smash{\mathbf{E}(\vec{k})}$. 

Once we have \eqsII{tricks_for_velocity_noise-A-2}{tricks_for_velocity_noise-A-3} we can rewrite the likelihood as 
\begin{equation}
\label{eq:tricks_for_velocity_noise-A-4}
\begin{split}
{\cal P}[\vec{\eps}_v] &= \int{\cal D}\mathbf{E}\,\Bigg(\prod_{\vec{k}}\frac{1}{(2\pi)^3}\Bigg) 
\exp\Bigg({-\frac{1}{2}}\int_{\vec{k}}P_vk^2\abs{\mathbf{E}(\vec{k})}^2\Bigg) 
\exp\bigg(\iu\int_{\vec{k}}\mathbf{E}(\vec{k})\cdot\vec{\eps}_v({-\vec{k}})\bigg)\,\,, 
\end{split}
\end{equation} 
which is also equal to 
\begin{equation}
\label{eq:tricks_for_velocity_noise-A-5}
\int{\cal D}\mathbf{E}\,\Bigg(\prod_{\vec{k}}\frac{1}{(2\pi)^3}\Bigg) 
\exp\bigg({-\frac{1}{2}}\int_{\vec{k}}P_vk^2\,\frac{\partial}{\partial\vec{\eps}_v(\vec{k})} 
\cdot\frac{\partial}{\partial\vec{\eps}_v({-\vec{k}})}\bigg) 
\exp\bigg(\iu\int_{\vec{k}}\mathbf{E}(\vec{k})\cdot\vec{\eps}_v({-\vec{k}})\bigg)\,\,. 
\end{equation}
Then, bringing the $\smash{\mathbf{E}}$-independent exponential out of the functional integral gives \eq{velocity_noise-14}.

%********************************************************
\subsection{Perturbative corrections to the EFT likelihood}
\label{app:tricks_for_velocity_noise-B}
%********************************************************

\noindent We now use \eq{velocity_noise-14} to sketch how to compute the corrections 
to the vanishing-noise likelihood perturbatively. First, we are integrating \eq{velocity_noise-12}, i.e.~ 
\begin{equation} 
\label{eq:tricks_for_velocity_noise-B-1} 
\Bigg(\prod_{\vec{x}}\sqrt{\frac{1}{2\pi P_{\eps_g}^{\{0\}}}}\,\Bigg) 
\exp\Bigg({-\frac{1}{2}}\int\dif^3x\,\frac{\big\{1 + 
\tilde{\delta}_g\big(\mathscr{R}^{-1}(\vec{x})\big) 
- \big(1+\delta_{g,{\rm det}}[\delta](\vec{x})\big)/J[\vec{v}_g](\vec{x})\big\}^2} 
{P_{\eps_g}^{\{0\}}/J^2[\vec{v}_g](\vec{x})}\Bigg)\,\,, 
\end{equation} 
multiplied by \eq{velocity_noise-14} over the velocity noise: we can then move the functional 
derivatives from the Dirac delta functional to \eq{velocity_noise-12}.\footnote{Since the derivatives 
always appear squared there is no need to keep track of signs.} Expanding the exponential to first order 
in $P_v$ we then need to compute, schematically, 
\begin{equation}
\label{eq:tricks_for_velocity_noise-B-2}
\bigg({-\frac{1}{2}}\int_{\vec{k}}P_vk^2\,\frac{\partial}{\partial\vec{\eps}_v(\vec{k})} 
\cdot\frac{\partial}{\partial\vec{\eps}_v({-\vec{k}})}\,\text{\eq{velocity_noise-12}}\bigg)\bigg|_{\vec{\eps}_v(\vec{k})=0}\,\,. 
\end{equation} 
Using the relation $\exp x\sim 1+x$, this can be then resummed into the logarithm of the likelihood. 

For simplicity, we consider only derivatives acting on the overall Jacobian $J$ (whose square multiplies 
the difference between theory and data in \eq{tricks_for_velocity_noise-B-1} above). 
When we expand the exponential in \eq{tricks_for_velocity_noise-B-1} 
in $\smash{1/P^{\{0\}}_{\eps_g}}$, and expand the Jacobian of the coordinate change in $\smash{\vec{\eps}_v}$ 
around the deterministic bias expansion for the galaxy velocity, 
we see that we have both ``disconnected'' contributions 
(that e.g.~modify the overall normalization) and ``connected'' contributions. 

We can see an example of the former already at first order in 
$\smash{1/P^{\{0\}}_{\eps_g}}$. We switch to Fourier space inside the exponential, and 
rewrite \eq{tricks_for_velocity_noise-B-1} as (we drop the overall factor for simplicity) 
\begin{equation} 
\label{eq:tricks_for_velocity_noise-B-3} 
\exp\bigg({-\frac{1}{2P_{\eps_g}^{\{0\}}}}\int_{\vec{p},\vec{p}'}\mathscr{J}(\vec{p})\mathscr{J}(\vec{p}') 
\mathscr{A}({-\vec{p}}-\vec{p}')\bigg)\,\,, 
\end{equation} 
where we defined the Fourier transform of the Jacobian as $\smash{\mathscr{J}}$, and the 
definition of $\smash{\mathscr{A}}$ can be seen by matching to \eq{tricks_for_velocity_noise-B-1}. 
The quantity to which we have to apply the differential operator of \eq{tricks_for_velocity_noise-B-2}, then, is 
\begin{equation}
\label{eq:tricks_for_velocity_noise-B-4}
{-\frac{1}{2P_{\eps_g}^{\{0\}}}}\int_{\vec{p},\vec{p}'}\mathscr{J}(\vec{p})\mathscr{J}(\vec{p}') 
\mathscr{A}({-\vec{p}}-\vec{p}')\,\,, 
\end{equation} 
where $\smash{\partial/\partial\vec{\eps}_v(\vec{k})}$ acts on $\smash{\mathscr{J}(\vec{p})}$ 
and $\smash{\partial/\partial\vec{\eps}_v({-\vec{k}})}$ on $\smash{\mathscr{J}(\vec{p}')}$. 
From \eqsII{notation-2}{velocity_noise-11} we have that 
\begin{equation}
\label{eq:tricks_for_velocity_noise-B-5}
\frac{\partial\mathscr{J}(\vec{k})}{\partial\vec{\eps}_v(\vec{k}')} = \frac{\iu\vec{k}\cdot\vers{n}}{{\cal H}}\,\vers{n}\,
(2\pi)^3\delta^{(3)}_{\rm D}(\vec{k}+\vec{k}')\,\,, 
\end{equation} 
so, using $\smash{\vers{n}\cdot\vers{n}=1}$, we find 
\begin{equation}
\label{eq:tricks_for_velocity_noise-B-6}
{-\frac{1}{4P_{\eps_g}^{\{0\}}}} 
\int_{\vec{k},\vec{p},\vec{p}'}P_vk^2\,\frac{(\vec{p}\cdot\vers{n})(\vec{p}'\cdot\vers{n})}{{\cal H}^2}\, 
(2\pi)^3\delta^{(3)}_{\rm D}(\vec{p}+\vec{k})\,(2\pi)^3\delta^{(3)}_{\rm D}(\vec{p}'-\vec{k})\,\mathscr{A}({-\vec{p}}-\vec{p}')\,\,. 
\end{equation} 
It is straightforward to see that this is proportional to $\smash{\mathscr{A}(\vec{0})}$. 

The connected contribution arises at second order in $\smash{1/P^{\{0\}}_{\eps_g}}$ 
from expanding the Jacobian that multiplies $\smash{1+\delta_{g,{\rm det}}[\delta](\vec{x})}$ 
in \eq{tricks_for_velocity_noise-B-1}. We have 
\begin{equation}
\label{eq:tricks_for_velocity_noise-B-8}
1 + \tilde{\delta}_g\big(\mathscr{R}^{-1}(\vec{x})\big) - \frac{1+\delta_{g,{\rm det}}[\delta](\vec{x})}{J[\vec{v}_g](\vec{x})}\supset 
\underbrace{1 + \tilde{\delta}_g\big(\mathscr{R}^{-1}(\vec{x})\big) 
- \frac{1 + \delta_{g,{\rm det}}[\delta](\vec{x})}{J[\delta,\!\vec{v}](\vec{x})}}_{
\hphantom{\mathscr{T}(\vec{x})}\equiv\,\mathscr{B}(\vec{x})} 
+ \frac{\vers{n}\cdot\partial_\parallel\vec{\eps}_v(\vec{x})}{\cal H}\,\,, 
\end{equation} 
where it is useful to assign a symbol, $\smash{\mathscr{B}(\vec{x})}$, to the square root of 
$\smash{\mathscr{A}(\vec{x})}$. Hence, losing track from now on of irrelevant overall factors 
and switching to Fourier space, we have to apply the differential operator of \eq{tricks_for_velocity_noise-B-2} to 
\begin{equation}
\label{eq:tricks_for_velocity_noise-B-9}
\int_{\vec{p}}\frac{(\vec{p}\cdot\vers{n})\,\vers{n}\cdot\vec{\eps}_v(\vec{p})}{{\cal H}}\, 
\mathscr{B}({-\vec{p}})\int_{\vec{q}}\frac{(\vec{q}\cdot\vers{n})\,\vers{n}\cdot\vec{\eps}_v(\vec{q})}{{\cal H}}\, 
\mathscr{B}({-\vec{q}})\,\,. 
\end{equation} 
After taking the derivatives with respect to the velocity noise, we arrive at 
\begin{equation}
\label{eq:tricks_for_velocity_noise-B-10}
\int_{\vec{k}}P_v k^2\frac{(\vec{k}\cdot\vers{n})^2}{{\cal H}^2}\mathscr{B}(\vec{k})\mathscr{B}({-\vec{k}})\sim 
{\frac{P_v}{{\cal H}^2}}\int\dif^3x\,\frac{\partial_\parallel\vec{\nabla}({\rm data}-{\rm theory})}{P_{\eps_g}^{\{0\}}} 
\cdot\frac{\partial_\parallel\vec{\nabla}({\rm data}-{\rm theory})}{P_{\eps_g}^{\{0\}}}\,\,, 
\end{equation} 
where we have schematically called ``${\rm data}-{\rm theory}$'' the Fourier transform back to real space of 
$\smash{\mathscr{B}}$ (it is straightforward to match with \eq{tricks_for_velocity_noise-B-1}, in 
the same way as when we defined $\smash{\mathscr{A}}$ and $\smash{\mathscr{B}}$ themselves). 
Moreover, the dimensions have been fixed by reinstating the overall factor of $\smash{1/P^{\{0\}}_{\eps_g}}$ 
squared that we had dropped throughout: $\smash{P_v/{\cal H}^2}$ has dimensions of a length to the $7$th power, 
$\partial_\parallel$ and $\vec{\nabla}$ have dimensions of an inverse length, $\smash{P^{\{0\}}_{\eps_g}}$ has 
dimensions of a length to the $3$rd power, and finally both $\smash{\dif^3x/P^{\{0\}}_{\eps_g}}$ and 
``${\rm data}-{\rm theory}$'' are dimensionless. 

We can then compare this with the logarithm of the likelihood for vanishing velocity noise. 
We immediately see that the presence of the square of ``$\smash{{\rm data}-{\rm theory}}$'' 
gives a scaling $\smash{\Lambda^3}$, c.f.~\eq{perturbativity-3}. 
The two derivatives $\smash{\vec{\nabla}}$ then give a scaling $\smash{\Lambda^2}$, and we recognize the overall $\smash{P_v}$. 
Importantly, we notice that there are two additional derivatives $\smash{\partial_\parallel}$ controlled 
by the Hubble scale $\smash{\cal H}$. This agrees with the conclusions of Section~\ref{subsec:checking_the_perturbative_expansion}. 
To derive the scaling with $\Lambda$, there we had used the fact that the ``typical scale of variation'' 
of functionals of $\smash{\vec{\eps}_v}$ was $\smash{\partial_\parallel/{\cal H}}$: 
\eqsII{tricks_for_velocity_noise-B-8}{tricks_for_velocity_noise-B-9} provide a clear example of this.

%********************************************************
%********************************************************
%********************************************************

%-------------------------------------------------------------------------------------------------------

\clearpage
%\cleardoublepage

\bibliographystyle{utphys}
\bibliography{refs}

%-------------------------------------------------------------------------------------------------------

\end{document}